\documentclass[aps,showpacs,onecolumn,notitlepage,floatfix,amsmath,amssymb,nofootinbib]{revtex4-1}

\usepackage{amssymb}
\usepackage{amsmath}
\usepackage{epsfig}
\usepackage{graphicx}
\usepackage{subfig}%%
\usepackage{color}
\usepackage{latexsym}
\usepackage{bm,latexsym}
\usepackage{mathrsfs}
\usepackage{float}
\usepackage{pbox}

\usepackage[usenames,dvipsnames]{xcolor}
\definecolor{orange}{cmyk}{0,0.5,1,0}

\begin{document}

\title{A novel exact magnetic black hole solution in four-dimensional extended scalar-tensor-Gauss-Bonnet theory}

\author{Pedro Ca\~nate$^{1}$}
\email{pcannate@gmail.com}

\author{Santiago Esteban Perez Bergliaffa$^{1}$}
\email{sepbergliaffa@gmail.com}

\affiliation{$^{1}$Departamento de F\'isica  Te\'orica, Instituto de F\'isica, Universidade do Estado do Rio de Janeiro,\\ Rua S\~ao Francisco Xavier 524, Maracan\~a \\
CEP 20550-013, Rio de Janeiro, Brazil.}

\begin{abstract}

In this work the first exact 
asymptotically flat static and spherically symmetric black hole solution for $(3+1)$-dimensional ESTGB is presented, with a model of nonlinear electrodynamics -that reduces to Maxwell's theory in the weak field limit and satisfies the weak energy condition- as a source. The solution has a nonzero magnetic charge, and scalar hair, which turns out to be dependent of the magnetic charge. 
It is characterized by the ADM mass $m$ and the magnetic charge $q$. Depending on the range of these parameters, the solution describes black holes with different structure. In the case $m\geq0$ and $q\geq0$, it shares many of the characteristics of the Schwarzschild solution. For $m>0$ and $q<0$, it is akin to the Reissner-Nordstr\"om metric. In the case $m=0$, it represents a purely magnetic black hole.

\end{abstract}

\pacs{04.20.Jb, 04.50.Kd, 04.50.-h, 04.40.Nr}

% 04.20.Jb Exact solutions
% 04.50.Kd Modified theories of gravity
% 04.50.-h Higher-dimensional gravity and other theories of gravity
% 04.40.Nr Einstein-Maxwell spacetimes, spacetimes with fluids, radiation or classical fields

\maketitle

\section{Introduction}

Many observational results and theoretical considerations
have led to the idea that Einstein's theory of General Relativity (GR) may not be valid in the high and/or low-curvature regimes. Among them we can mention the present stage of cosmic acceleration
(see \cite{Yang2019} and references 
therein), the flatness of the rotation curves of galaxies
(see for instance \cite{Capo2006}), 
the union of gravity with the laws of quantum physics
(see for instance \cite{Fradkin1985}), 
and 
current limits on the tensor-to-scalar ratio \cite{Akrami2018}.
%%%%%%%%%%%%%%%%%%%%%%%%%%%%%%%%%%%%%%%%
Most of the extensions of GR involve 
additional degrees of freedom, such as those coming from a scalar field and/or 
curvature corrections to the  Einstein-Hilbert Lagrangian, coupled to
the scalar field. A particular and interesting extension, since it avoids the Ostrogradski instability,
is given by the coupling of the 
scalar field with the
Gauss-Bonnet invariant, which yields the so-called extended scalar-tensor-Gauss-Bonnet (ESTGB) 
(see \cite{scalariz_Doneva} for details)
\footnote{In an $N$-dimensional setting
and vanishing scalar field, the resultant theory, known as Einstein-Gauss-Bonnet gravity has very interesting properties, 
such as 
regular cosmological solutions 
(in $4+1$ dimensions), 
\cite{Cosm_GB}, and  $(n+1)$-dimensional  black hole solutions (with $n+1>4$), that have no GR limit
\cite{Cai, Anabalon}.}.
Such a theory follows 
at the level of the effective action
of string theory, when viewed in the Einstein frame \cite{eff}.

Various aspects of the ESTGB theory
have been studied in detail lately. To name just a few applications in cosmology, 
it was shown in 
\cite{cosmic_acceleration} that 
the theory can describe 
the
present stage of cosmic acceleration, and
can lead to an exit from a
scaling matter-dominated epoch to a late-time accelerated expansion \cite{Tsujikawa2006}. The possible reconstruction of the coupling and potential functions for a given scale factor was considered in \cite{recon}, and
the consequences of ESTGB in an inflationary setting have been considered in \cite{Odintsov2020}.

Black hole solutions have been extensively discussed in ESTGB in $(3+1)$-dimensions, without matter fields. Given the complexity of the field equations, at present there are
only numerical solutions available. In particular, and restricting to the static and spherically symmetric case, 
new black hole were shown to form 
by spontaneous scalarization -induced by the curvature of the geometry- of the Schwarzschild black holes
in the extreme curvature regime
in \cite{Doneva2018, scalariz_Silva}
\footnote{The linear stability of such configurations was studied in \cite{Kunz2018}, and the polar quasinormal modes, in
\cite{Blazquez2020}.}
(see \cite{Doneva2019} for a case of a massive scalar field, \cite{scalariz_Doneva} for the charged case, 
\cite{kanti96} for the dilatonic case, 
and \cite{Doneva2020} for the multi-scalar case). 
There are also numerical solutions displaying ``natural scalarization'', such as those presented in  \cite{Kanti2018,Ultra_compact,Antoniou18}). Since all the solutions found in the literature are numerical, it would be of interest to find 
analytical solutions. We would like to present here the first solution of such a kind, 
using nonlinear electrodynamics (NLED) as a source. 

Maxwell's electrodynamics is a very well-established theory  which
has been subjected to innumerable tests. However, 
there are good theoretical reasons to consider modifications to it. For instance, 
with the aim of avoiding the divergent behavior associated to the electric field
and the self-energy of a point charge, 
M. Born and L. Infeld \cite{NLED} proposed a modified version of Maxwell's theory that was a nonlinear function of the two
electromagnetic invariants, 
namely
$\mathcal{F} = 2(\mathcal{B}^{2}-\mathcal{E}^{2})$ and $\mathcal{G} = \mathcal{E}\!\cdot\!\mathcal{B}$.

The explicit form of the Lagrangian $\mathcal{L}(\mathcal{F},\mathcal{G})$ may be determined according 
to different criteria 
\footnote{See \cite{genNLED} for a general framework 
for NLED.}, but there are two examples 
that are particularly relevant: 
that of the Born-Infeld theory mentioned above\footnote{For a modern take on the BI action see for instance
\cite{Fradkin1985b, Gibbons2000}
.}, and the Euler-Heisenberg theory \cite{Euler_Heisenberg}, derived from one-loop quantum electrodynamics, that 
describes some nonlinear processes, like the light-by-light scattering. 
%%%%%%%%%%%%%%%%%%%%%%%%%%%%%%%%%%%%%%%%
The latter 
is a quantum-mechanical process that is forbidden in classical 
electrodynamics. This reaction is accessible at the Large Hadron Collider thanks to the large electromagnetic field strengths generated by ultra-relativistic colliding lead ions (see for details \cite{lightbylight}, where experimental evidence for light-by-light scattering has been reported).

It can also be added that the coupling of GR with NLED has led to a number of interesting solutions and phenomena. Among them we can mention black holes with everywhere-regular  curvature invariants and electric field (see for instance \cite{BH_NLED}); traversables wormholes sustained with nonlinear electromagnetic fields \cite{WH_NLED}; non-gravitational wormholes \cite{Baldovin2000}, and a new entropy bound \cite{Falciano2019}. 

In this article, the first exact black hole solution derived for $(3+1)-$dimensional extended scalar-tensor-Gauss-Bonnet theory (ESTGB) coupled to a particular form of nonlinear electrodynamics (NLED) is presented
\footnote{The inclusion of electromagnetic fields in ESTGB theory has been recently implemented in order to construct traversable wormhole (T-WH) exact solutions \cite{Exact_wH_ED,Exact_wH_NLED}, that do not require exotic matter, in where the scalar-Gauss-Bonnet curvature is the only responsible for the negative energy density
necessary for the traversability.
}. 
The electromagnetic Lagrangian is such that it reduces to that of Maxwell in the weak field limit. 
The solution is analyzed in detail, and it is shown that, depending on the parameters that determine the spacetime metric, black holes with different features are 
allowed. 
The paper is organized as follows: in the next section we briefly outline the field equations derived from the ESTGB-NLED action. In Sect. \ref{two_parametric} the derived  two-parametric family of solutions is presented,
its black hole interpretation is analyzed, are treated the limiting cases of vanishing electromagnetic and scalar field, and vanishing of the ADM mass. Final conclusions are given in the last section.
In this paper we use units where $G = c = 1$.

%%%%%%%%%%%%%%%%%%%%%%%%%%%%%%%%%%%%%%%%%%%%%%%%%%%%%%%%%%%%%%%%%%%%%%%%%%%%%%%%%%%%%%%%%%%%%%%%%%%%%%%%%%%%%%
\section{ Extended Scalar-Tensor-Gauss-Bonnet gravity } 

The $(3+1)$-dimensional extended scalar-tensor-Gauss-Bonnet theory with additional matter fields, is defined by the following action,
%%%%%%%%%%%%%%%%%%%%%%%%%%%%%%%%%%%%
\begin{equation}\label{actionL}
S[g_{ab},\phi,\psi_{a}] = \int d^{4}x \sqrt{-g} \left\{ \frac{1}{16\pi}\left(R - \frac{1}{2}\partial_{\mu}\phi\partial^{\mu}\phi + \boldsymbol{f}(\phi) R_{_{GB}}^{2} - 2 \mathscr{U}(\phi) \right) - \frac{1}{4\pi}\mathcal{L}_{\rm matter}(g_{ab},\psi_{a})  \right\}.
\end{equation}
The first term in this action defines the Einstein-Hilbert Lagrangian density. The Lagrangian density for the STGB contribution
is defined by the addition of the kinetic term of the scalar field, the quadratic Gauss-Bonnet term
$R_{_{GB}}^{2}$
non-minimally coupled to the scalar field, by the function
$\boldsymbol{f}(\phi)$, and the scalar field potential $\mathscr{U}(\phi)$. The Lagrangian density $\mathcal{L}_{\rm matter}(g_{ab},\psi_{a})$
represents any matter fields present in the system. In this work, we will consider the 
ESTGB theory in the presence of 
non-linear electrodynamics (NLED), in which the Lagrangian  
is a function on the electromagnetic invariant $\mathcal{F}=F_{\alpha\beta}F^{\alpha\beta}/4$, with $F_{ab}=2\partial_{[a}A_{b]}$, and $A_{a}$ the electromagnetic potential.
%%%%%%%%%%%%%%%%%%
The ESTGB-NLED field equations arising from
the variation of the action with respect to  
$g_{ab}$ are 
\begin{equation}\label{modifEqs} 
G_{a}{}^{b} = 8\pi (E_{a}{}^{b})\!_{_{_{G\!B}}} + 8\pi (E_{a}{}^{b})\!_{_{_{N \! L \! E \! D}}}, 
\end{equation}
where $G_{a}{}^{b} = R_{a}{}^{b} - \frac{R}{2} \delta_{a}{}^{b}$ denotes the components of the Einstein tensor,  
and the quantities $(E_{a}{}^{b})\!_{_{_{G\!B}}}$ and $(E_{a}{}^{b})\!_{_{_{N \! L \! E \! D}}}$ are defined by the following expressions:
%%%%%%%%%%%%%%%%%%%
\begin{eqnarray}
&& 8\pi (E_{\alpha}{}^{\beta})\!_{_{_{G\!B}}} = -\frac{1}{4}(\partial_{\mu}\phi \partial^{\mu}\phi)\delta_{\alpha}{}^{\beta} + \frac{1}{2}\partial_{\alpha}\phi \partial^{\beta}\phi - \frac{1}{2}( g_{\alpha\rho} \delta_{\lambda}{}^{\beta} + g_{\alpha\lambda} \delta_{\rho}{}^{\beta})\eta^{\mu\lambda\nu\sigma}\tilde{R}^{\rho\xi}{}_{\nu\sigma}\nabla_{\xi}\partial_{\mu}\boldsymbol{f}(\phi)  - \mathscr{U}(\phi)\delta_{\alpha}{}^{\beta} \label{E_GB} ,\\
&&8\pi (E_{\alpha}{}^{\beta})\!_{_{_{N \! L \! E \! D}}} =  2\left( \mathcal{L}_{\mathcal{F}} \hskip.02cm F_{\alpha\mu}F^{\beta\mu} - \mathcal{L}\hskip.03cm\delta_{\alpha}{}^{\beta}\right), \label{E_NLED}
\end{eqnarray}
with $\tilde{R}^{\rho\gamma}{}_{\mu\nu} = \eta^{\rho\gamma\sigma\tau}R_{\sigma\tau\mu\nu} = \epsilon^{\rho\gamma\sigma\tau}R_{\sigma\tau\mu\nu}/\sqrt{-g}$, and $\mathcal{L}_{\mathcal{F}}\equiv \frac{d\mathcal{L}}{d\mathcal{F}}$ .
Thus, $(E_{\alpha}{}^{\beta})\!_{_{_{G\!B}}}$ denotes the components of a tensor which we shall refer to as the Scalar-Gauss-Bonnet tensor, since it represents
the contribution to the spacetime curvature due to the effects of the STGB term. The components of the NLED energy-momentum tensor
are denoted by
$(E_{\alpha}{}^{\beta})\!_{_{_{N \! L \! E \! D}}}$. The structure of the field equations (\ref{modifEqs})
motives the definition of the effective energy-momentum tensor, $\mathscr{E}_{a}{}^{b}$, as $\mathscr{E}_{a}{}^{b} = (E_{a}{}^{b})\!_{_{_{G\!B}}} +  (E_{a}{}^{b})\!_{_{_{N \! L \! E \! D}}}$. Thus, the ESTGB-NLED theory can be written 
in a GR-like form, 
$G_{a}{}^{b} = 8\pi\mathscr{E}_{a}{}^{b}$.
%%%%%%%%%%%%%%%%%%%%%%%%%%%%%%%%%%%%%%%%%%%%%%%%%%%%%%%%%%%%%%%%%%%%%%%%%%%%%%%%%%%%%%

By taking the divergence of Eq.(\ref{modifEqs}),
and taking into account that 
$\nabla_\beta (E_{\alpha}{}^{\beta})\!_{_{_{N\! L \! E \! D}}}=0$, and that the Bianchi identities guarantee that $\nabla_{\beta} G_{\alpha}{}^{\beta}=0$, it follows that 
$\nabla_\beta(E_{\alpha}{}^{\beta})\!_{_{_{G\!B}}}=0$. This in turn implies that 
%
%%%%%%%%%%%%%%%%%%%%%%%%%%%%%%%%
\begin{equation}\label{scalar_Eq}
\nabla^{2}\phi + \dot{\boldsymbol{f}}(\phi) R_{_{GB}}^{2} + 2 \dot{\mathscr{U}}(\phi) = 0,
\end{equation}
where the overdot 
denotes the derivatives with respect to the scalar
field, i.e., ($\dot{\boldsymbol{f}} = \frac{d\boldsymbol{f}}{d\phi}$, $\dot{\mathscr{U}} = \frac{d\mathscr{U}}{d\phi}$). 

The equations of motion for the nonlinear electrodynamical theory are given by
\begin{equation}\label{EM_Eqs}
\nabla_{\alpha}( \mathcal{L}_{\mathcal{F}}F^{\alpha\beta} ) = 0,
\;\;\;\;\;\;\;\;\;\; \nabla_{\alpha}( _{\ast}\!\boldsymbol{F} )^{\alpha}=0, \end{equation}
where $_{\ast}\!\boldsymbol{F}$ denotes the Hodge star operation (or Hodge dual) with respect to the metric.
Our aim is to find a exact solution of the set of Eqs. (\ref{modifEqs}), (\ref{scalar_Eq}), and (\ref{EM_Eqs}),
that describe an asymptotically flat, static and spherically symmetric black hole (AF-SSS-BH). 
Therefore, we will assume that the scalar field is static and spherically symmetric, $\phi = \phi(r)$, the electromagnetic invariant depends only on the radial coordinate, $\mathcal{F}=\mathcal{F}(r)$, and also that the metric takes the static and spherically symmetric form,
\begin{equation}\label{SSSmet}  
ds^{2} =  - e^{ A(r) }dt^{2} + e^{ B(r) }dr^{2}  + r^{2}(d\theta^{2}  + \sin^{2}\theta d\varphi^{2}),
\end{equation}
with $A = A(r)$ and $B = B(r)$ unknown functions 
of $r$. 

Regarding the electromagnetic field tensor, since the spacetime is static and spherically symmetric, then the only non-vanishing terms are the electric component $F_{tr} = \mathcal{E}(r)$ and the magnetic component $F_{\theta\varphi} = \mathcal{B}(r,\theta)=h(r)\sin\theta$. In this work we restrict ourselves to a purely magnetic field, i.e., $\mathcal{E}(r) = 0$ and $\mathcal{B}(r,\theta) \neq 0$. 
In this way, for a static and spherically symmetric spacetime with line element (\ref{SSSmet}), the general solution of Eq. (\ref{EM_Eqs}) is given by,
%%%%%%%%%%%%%%%%%%%%%%%%%%%%%%%%%%%%%
\begin{equation}\label{fabSOL}
F_{\theta\varphi} =  r^{4} \mathcal{Q}(r) \sin\theta  .
\end{equation}
Then, $\boldsymbol{F} = r^{4} \mathcal{Q}(r) d\theta \wedge d\varphi$, therefore $d\boldsymbol{F} = 0 = (r^{4} \mathcal{Q}(r))' dr \wedge d\theta \wedge d\varphi$, (where the prime denotes the derivative with respect to the radial coordinate $r$) yields $\mathcal{Q}(r) = \sqrt{2}\hskip.06cm q/r^{4}$, where
$\sqrt{2}\hskip.06cm q$ is an integration constant,  which it plays the role of the magnetic charge. 
Hence, the 
components of the electromagnetic field tensor, and the invariant $\mathcal{F}$ are respectively given by; 
%%%%%%%%%%%%%%%%%%%%%%%%%%%%%%%%%%%%%%%
\begin{equation}\label{magnetica}
F_{\alpha\beta} = F_{\theta\varphi} (\delta_{\alpha}^{\theta}\delta_{\beta}^{\varphi} - \delta_{\alpha}^{\varphi}\delta_{\beta}^{\theta}),\quad\quad\quad\quad\quad F_{\theta\varphi}= \mathcal{B}(\theta) = \sqrt{2}\hskip.06cm q\sin\theta, \quad\quad\quad\quad\quad \mathcal{F} = \frac{ q^{2} }{ r^{4} }.
\end{equation}
%%%%%%%%%%%%%%%%%%%%%%%%%
% 
%%%%%%%%%%%%%%%%%%%%%%%%%

\section{Field equations}

For the line element (\ref{SSSmet}), the non-null 
components of the
Einstein tensor are given by
%%%%%%%%%%%%%%%%%%%%%%%%%
\begin{eqnarray}
G_{t}{}^{t}\!=\!\frac{ e^{^{\!\!-B}} \left(\! -rB' - e^{^{\!B}} + 1 \!\right) }{ r^{2} }\!, \hspace{0.19cm}  G_{r}{}^{r} \!=\! \frac{ e^{^{\!\!-B}} \left(\! rA' - e^{^{\!B}} + 1 \!\right) }{ r^{2} }\!, \hspace{0.19cm} G_{\theta}{}^{\theta}\!=\! G_{\varphi}{}^{\varphi}\!=\!\frac{ e^{^{\!\!-B}} \left( rA'^{2} - rA'B' + 2rA'' + 2A' - 2B' \right) }{ 4r }\!.
\end{eqnarray}
%%%%%%%%%%%%%%%%%%%%%%%%%
%%%%%%%%%%%%%%%%%%%%%%%%%%
The non-vanishing 
components of the 
SGB tensor 
with arbitrary coupling function $\boldsymbol{f}(\phi)$ and potential
$\mathscr{U}$
are
%%%%%%%%%%%%%%%%%%%%%%%%%
\begin{eqnarray}
&& 8\pi(E_{t}{}^{t})\!_{_{_{G\!B}}} = -\frac{ e^{ -2B} }{ 4r^{2} }\left\{  \left[r^{2}e^{B} + 16(e^{B} - 1)\ddot{\boldsymbol{f}} \right]\phi'^{2} - 8[ (e^{B} - 3)B'\phi' - 2(e^{B} - 1)\phi'']\dot{\boldsymbol{f}} \right\} - \mathscr{U}, \label{GBtt} \\
&& 8\pi(E_{r}{}^{r})\!_{_{_{G\! B}}} = \frac{ e^{ -B} \phi'}{ 4 } \left[ \phi'  - \frac{8(e^{B} - 3)e^{-B}A'\dot{\boldsymbol{f}} }{r^{2}}  \right] - \mathscr{U},  \label{GBrr} \\
&& 8\pi(E_{\theta}{}^{\theta})\!_{_{_{G\! B}}} = (E_{\varphi}{}^{\varphi})\!_{_{_{G\! B}}} = -\frac{ e^{ -2B} }{ 4r } \left\{ ( re^{B} - 8A'\ddot{\boldsymbol{f}})\phi'^{2} - 4\left[ (A'^{2} + 2A'')\phi' + (2\phi'' - 3B'\phi')A'\right]\dot{\boldsymbol{f}} \right\} - \mathscr{U}.
\end{eqnarray}
Finally, the  energy-momentum tensor components 
for NLED, 
assuming the SSS spacetime with metric (\ref{SSSmet}), the electromagnetic field tensor (\ref{magnetica}), and a Lagrangian density $\mathcal{L}(F)$, are given by 
%%%%%%%%%%%%%%%%%%%%%%%%%
\begin{eqnarray}\label{E_nled}
8\pi (E_{t}{}^{t})\!_{_{_{N \! L \! E \! D}}} = 8\pi (E_{r}{}^{r})\!_{_{_{N \! L \! E \! D}}} =  -2\mathcal{L},   
\quad\quad\quad 8\pi (E_{\theta}{}^{\theta})\!_{_{_{N \! L \! E \! D}}} = 8\pi (E_{\varphi}{}^{\varphi})\!_{_{_{N \! L \! E \! D}}} =  2(2\mathcal{F}\mathcal{L}_{\mathcal{F}} - \mathcal{L}). 
\end{eqnarray}
%%%%%%%%%%%%%%%%%
%%%%%%%%%%%%%%%%%%%%%%%%%
%
%%%%%%%%%%%%%%%%%%%%%%%%%
Inserting the above given components in the field equations (\ref{modifEqs}), 
we obtain:
%%%%%%%%%%%%%%%%%%%%%%%%%
\begin{eqnarray}
&&\!G_{t}{}^{t}\!=\!8\pi\mathscr{E}_{t}{}^{t}\!\hskip.2cm\Rightarrow\hskip.2cm 
4e^{B}\!\!\left( rB' \!+\! e^{ B} \!-\! 1 \right) \!=\!\!\left[ r^{2}e^{B}\!+\!16(e^{ B}\!-\!1)\ddot{\boldsymbol{f}} \right]\!\!\phi'^{2} \!-\!8\!\left[ (e^{ B} \!-\!3)B'\phi' \!-\! 2(e^{ B} \!-\! 1)\phi'' \right]\!\!\dot{\boldsymbol{f}} 
\!+\! 4r^{2}e^{2B}(\mathscr{U}\!+\!2\mathcal{L}), \label{Eqt}\\
&&\nonumber\\
&&\!G_{r}{}^{r}\!=\!8\pi\mathscr{E}_{r}{}^{r}\!\hskip.115cm\Rightarrow\hskip.115cm 4e^{B}\!\!\left( -rA'\!+\! e^{ B} \!-\! 1 \right) \!=\! - r^{2}e^{B}\phi'^{2} \!+\! 8(e^{ B} \!-\! 3)A'\phi'\dot{\boldsymbol{f}} \!+\! 4r^{2}e^{2B}(\mathscr{U}\!+\!2\mathcal{L}),
\label{Eqr}\\
&&\nonumber\\
&&\!G_{\theta}{}^{\theta}\!=\!8\pi\mathscr{E}_{\theta}{}^{\theta}\!\hskip.115cm\Rightarrow\hskip.115cm e^{B}\!\!\left\{ rA'^{2} \!-\! 2B' \!+\! (2 \!-\! rB')A' \!+\! 2rA'' \right\} \!=\! -re^{B}\phi'^{2} \!+\! 8A'\ddot{\boldsymbol{f}}\phi'^{2} \nonumber \\
&& \hskip6.9cm \!+ 4\!\left[ (A'^{2} \!+\! 2A'')\phi' \!+\! (2\phi'' \!-\! 3B'\phi')A' \right]\!\!\dot{\boldsymbol{f}} \!-\! 4 r e^{2B}(\mathscr{U}\!+\!2\mathcal{L}\!-\!4\mathcal{F}\mathcal{L}_{\mathcal{F}}).\label{Eqte}
\end{eqnarray}
The equation 
(\ref{scalar_Eq})
for the scalar field $\phi$ can be written as
%%%%%%%%%%%%%%%%%%%%%
\begin{equation}\label{phi2}
2r\phi'' + (4 + rA' - rB')\phi' + \frac{4e^{-B}\dot{\boldsymbol{f}}}{r} \left[ (e^{B} - 3)A'B' - (e^{B} - 1)(2A'' + A'^{2})\right] - 4r e^{B}\dot{\mathscr{U}} = 0.
\end{equation}
In the case with $\mathscr{U}(\phi)$=$\Lambda$=constant, and $\mathcal{L}(\mathcal{F})$=$0$, the system of equations (\ref{Eqt}), (\ref{Eqr}), (\ref{Eqte}) and (\ref{phi2}), reduces to that for EGB gravity with a nonminimally coupled massless scalar field in the presence of a cosmological constant, see for instance \cite{Kanti2018}.
%%%%%%%%%%%%%%%%%%%%%

The problem of describing an AF-SSS magnetic black hole solution within the framework of ESTGB-NLED gravity, reduces to solving the field equations (\ref{Eqt}), (\ref{Eqr}), (\ref{Eqte}) and (\ref{phi2}), with electromagnetic field (\ref{magnetica}), and SSS metric (\ref{SSSmet}). It is useful to introduce the functions 
$\mathscr{M}$ and $\delta$ through
%%%%%%%%%%%%%%%%%%%%%
\begin{equation}\label{SSS_ansatz}
e^{A(r)} = \left( 1 - \frac{2\mathscr{M}\!(r)}{r}\right)e^{2\delta(r)}, \quad\quad\quad\quad e^{B(r)} = \left( 1 - \frac{2\mathscr{M}\!(r)}{r}\right)^{-1},    \end{equation}
where the mass function $\mathscr{M}\!(r)$ provides the ADM (Arnowitt-Deser-Misner) mass ($\mathscr{M}_{_{\!\!A\!D\!M}}$)  in the asymptotic region, $\mathscr{M}_{_{\!\!A\!D\!M}} = \lim\limits_{r\rightarrow \infty}\mathscr{M}\!(r)\in\mathbb{R}^{+}$, and $\lim\limits_{r\rightarrow \infty}\delta(r) = 0$, provided the spacetime is AF. The behavior of the scalar field 
$\phi$ in the limit $r\rightarrow\infty$
 must be
\begin{equation}
\label{scalar}
\phi(r) = \phi_{0} + \phi_{1}/r + \phi_{2}/r^{2} + \mathcal{O}\!\left(r^{-3}\right), 
\end{equation}
with $\phi_{i}$ real parameters, and $\phi_{1}$ the scalar charge, see  \cite{kanti96,Antoniou18,Kanti2018,Ultra_compact} for details. 

Regarding the NLED Lagrangian
$\mathcal{L}(\mathcal{F})$, we shall impose, 
in agreement with Born and Infeld \cite{NLED}, 
that it reduces to
the Maxwell
Lagrangian in the 
weak-field limit, i.e., $\mathcal{L} \rightarrow \kappa \hskip.05cm\mathcal{F}$, $\mathcal{L}_{\mathcal{F}} \rightarrow \kappa$ (being $\kappa$ a constant), when $\mathcal{F}$
is very small.
%%%%%%%%%%%%%%%%%%%%%%%%%%%%%%%%%%%%%%%%%%%%%%%%%%%%%%%%%%%%%%%%%%%%%%%%%%%%%%%%%%%%%%%%%%%%%%%%%%%%%%%%%%%%%%%%%%%%
%  
%%%%%%%%%%%%%%%%%%%%%%%%%%%%%%%%%%%%%%%%%%%%%%%%%%%%%%%%%%%%%%%%%%%%%%%%%%%%%%%%%%%%%%%%%%%%%%%%%%%%%%%%%%%%%%%%%%%%
Furthermore, 
we shall require  that the corresponding NLED energy-momentum tensor $(E_{a}{}^{b})\!_{_{_{N \! L \! E \! D}}}$ satisfies the  weak energy condition (WEC), which states that for any timelike vector $\boldsymbol{k} = k^{\mu}\partial_{\mu}$, (i.e., $k_{\mu}k^{\mu}<0$), the tensor $(E_{a}{}^{b})\!_{_{_{N \! L \! E \! D}}}$ obeys the inequality  
$(E_{\mu\nu})\!_{_{_{N \! L \! E \! D}}}k^{\mu}k^{\nu} \geq 0$, which means that the local energy density $\rho_{\!_{_{loc}}}=(E_{\mu\nu})\!_{_{_{N \! L \! E \! D}}}k^{\mu}k^{\nu}$ as measured by any observer with timelike vector $\boldsymbol{k}$ is a non-negative quantity.
Following \cite{WEC}, for a diagonal energy-momentum tensor $(E_{\alpha\beta})=diag \left( E_{tt},E_{rr},E_{\theta\theta},E_{\varphi\varphi} \right)$, which can conveniently be written as
%%%%%%%%%%%%%%%%%%%%%%%%%%%%%%%%%%%%%
\begin{equation}\label{diagonalEab}
E_{\alpha}{}^{\beta} = - \rho \hskip.05cm \delta_{\alpha}{}^{t}\delta_{t}{}^{\beta} + P_{r} \hskip.05cm \delta_{\alpha}{}^{r}\delta_{r}{}^{\beta} + P_{\theta} \hskip.05cm \delta_{\alpha}{}^{\theta}\delta_{\theta}{}^{\beta} + P_{\varphi} \hskip.05cm \delta_{\alpha}{}^{\varphi}\delta_{\varphi}{}^{\beta} ,   
\end{equation}
WEC leads to
\begin{equation}\label{WEC}
\rho = - E_{t}{}^{t} \geq0, \quad \rho + P_{a} \geq0, \quad  a = \{ r, \theta, \varphi \}.     
\end{equation}
%%%%%%%%%%%%%%%%%%%%%%%%%%%%%%%
  
\section{An exact two-parametric family of
 solutions in ESTGB theory}\label{two_parametric}
 
Let us present now the specific choice of the relevant functions 
that leads to an exact static and spherically symmetric solution. 
$\boldsymbol{f}(\phi)$, $\mathscr{U}(\phi)$, and 
$\mathcal{L}$ are respectively
given by 
%%%%%%%%%%%%%%%%%%%%%%%%%%%%%%%%%%%%
%%%%%%%%%%%%%%%%%%%%%%%%%%%%%%%%%%%%
\begin{eqnarray} 
\boldsymbol{f}\!&=&\!-\frac{\ell^{2}\sigma}{32}\!\!\left\{ \sqrt{2\sigma}\tan^{\!^{\!-1}}\!\!\!\left( \frac{\sqrt{2} }{ \sqrt{\sigma}\hskip.06cm \phi}\right) +\frac{1}{2\phi}\ln\!\!\left[\!\left(\!\!\frac{ 2\beta }{ \sigma \phi^{2} } \!+\! \beta\!\!\right)^{\!\!\!2} \right] - \frac{2}{\phi} \right\}\!\!, 
\label{fNew}\\%%%%%%%%
&&\nonumber\\
\mathscr{U}\!\!&=&\!\frac{ 2^{\!^{\frac{9}{2}}}}{105\ell^{2}\sigma^{\frac{7}{2}}}\!\left[\frac{\pi}{2}-\!\tan^{\!^{\!-1}}\!\!\!\left(\!\!\frac{\sqrt{2} }{ \sqrt{\sigma}\hskip.06cm \phi}\!\!\right)\right]\!+\!\frac{\phi^{5}}{4\ell^{2}}\!\!\left(\!\frac{3}{10\sigma}\!+\!\frac{5\phi^{2}}{7}\!+\!\frac{7\sigma\phi^{4}}{24} \!\right)\!\!\ln\!\!\left[\!\!\left(\!\!\frac{ 2\beta }{ \sigma \phi^{2} } \!+\! \beta \!\!\right)^{\!\!\!2}\right]\!-\! \frac{\phi}{3\ell^{2}}\!\!\left(\!\frac{16}{35\sigma^{3}} \!-\! \frac{8\phi^{2}}{105\sigma^{2}}\!+\! \frac{31\phi^{4}}{70\sigma}\!+\! \frac{11\phi^{6}}{28}\!\right)\!\!, \label{UNew} 
\end{eqnarray}
\begin{eqnarray}\label{NLEDt}
%%%%%%%%
\mathcal{L}\!&=&\!\frac{\mathcal{F}}{8}\!+\!\frac{8\mathcal{F}^{\frac{1}{4}}}{105 \sigma_{\!\ast}^{3} s^{\frac{3}{2}}  }\!-\!\frac{4\mathcal{F}^{\frac{3}{4}}}{315 \sigma_{\!\ast}^{2} s^{\frac{1}{2}}}\!-\!\left(\!\frac{37}{210\sigma_{\!\ast}}\!+\!1\!\right)\!s^{\frac{1}{2}}\mathcal{F}^{\frac{5}{4}} \!-\!\frac{5s^{\frac{3}{2}}\mathcal{F}^{\frac{7}{4}}}{84}\!+\!\frac{ 2^{ \!^{\frac{7}{2}} } }{ 105 \sigma_{\!\ast}^{\frac{7}{2}} s^{2} }\!\left[\!\tan^{\!^{\!-1}}\!\!\!\left(\!\!\frac{\sqrt{2} }{ \sqrt{ s\sigma_{\!\ast} }\hskip.06cm \mathcal{F}^{\frac{1}{4}} }\!\!\right) - \frac{\pi}{2}\right] \nonumber \\
&&\hskip5.4cm-\frac{1}{2}\!\!\left(\!\frac{\mathcal{F}}{8}\!-\!\frac{3s^{\frac{1}{2}}\mathcal{F}^{\frac{5}{4}}}{10\sigma_{\!\ast}}\!+\!\frac{3\sigma_{\!\ast} s\mathcal{F}^{\frac{6}{4}}}{16}\!-\!\frac{4 s^{\frac{3}{2}}\mathcal{F}^{\frac{7}{4}} }{7}\!-\!\frac{ 5\sigma_{\!\ast} s^{ \frac{5}{2} } \mathcal{F}^{\frac{9}{4}} }{24}\!\right)\!\ln\!\!\left[\!\left(\!\frac{2\beta_{\ast}}{ \sigma_{\!\ast} s \mathcal{F}^{\frac{1}{2}} }\!+\!\beta_{\ast}\!\right)^{\!\!2}\right]\!. %-\!\frac{ 2^{\!^{\frac{5}{2}}} \pi}{105s^{2}\sigma^{\frac{7}{2}}_{\!\ast} }. 
\end{eqnarray}
The
constants 
$\ell$, $\sigma$, and $\beta$ belong to the ESTGB sector, while $s$, $\sigma_{\!\ast}$ and $\beta_{\ast}$, are in the NLED sector. 
%%%%%%%%%%%%%%%%%%%%%%%%%%%%%%%%%%%%%%%%%%%%%%%%%%%%%%%%%%%%%%%%%%%%%%%%%%%%%%%%%%%%%%%%%%%%%%%%%%%%%%%%%%%%%%%%%%%%

In order that the NLED Lagrangian shown in 
Eq.\eqref{NLEDt}
reduces in the weak-field to Maxwell's electrodynamics,
we will assume that 
such
Lagrangian 
is valid for $\mathcal{F}
> \mathcal{F}_{_{0}}=\frac{ 4\beta^{2}_{\ast} }{ s^{2}\sigma^{2}_{\!\ast} }$. 
Notice that $\mathcal{F}_{_{0}}$ can in principle be made as small as needed, by choosing the appropriate value of the constants in its definition.
This minimum value guarantees the correct weak-field limit, 
as follows from the 
expansion of Eq.(\ref{NLEDt}) for small $\mathcal{F}$:
%  
%%%%%%%%%%%%%%%%%%%%%%%%%%%%%%%%%%%%%%%%%%
\begin{eqnarray}\label{Lder}
&&\mathcal{L} \!=\!\frac{1}{8}\!\left[\!1\!-\!\frac{1}{2}\!\ln\!\!\left(\!\frac{4\beta_{\ast}^{2}}{ \sigma_{\!\ast}^{2} s^{2}  \mathcal{F} }\!\right)\! \right]\!\!\mathcal{F}\!-\!s^{^{\frac{1}{2}}}\!\!\!\left[ 1 \!+\! \frac{37}{210\sigma_{\!\ast}} \!+\! \frac{2}{525\sigma_{\!\ast}}\!-\! \frac{3 }{20\sigma_{\!\ast}}\ln\!\!\left(\!\frac{4\beta_{\ast}^{2}}{  \sigma_{\!\ast}^{2} s^{2} \mathcal{F}}\!\right)\!\right]\!\!\mathcal{F}^{\frac{5}{4}}\!-\!\frac{ \sigma_{\!\ast} s }{16}\!\left[\!1\!+\frac{3}{2}\ln\!\!\left(\!\frac{4\beta_{\ast}^{2}}{  \sigma_{\!\ast}^{2} s^{2} \mathcal{F} }\!\right) \right]\!\!\mathcal{F}^{\frac{3}{2}}\!\nonumber\\
&&\hskip9cm +\frac{ s^{^{\frac{3}{2}}} }{7}\!\!\left[\!\frac{9}{14}\!+\!2\ln\!\!\left(\!\frac{4\beta_{\ast}^{2}}{  \sigma_{\!\ast}^{2} s^{2} \mathcal{F} }\!\right)\!\right]\!\!\mathcal{F}^{\frac{7}{4}}\!-\!\frac{5s^{2}\sigma^{2}_{\!\ast}}{64}\!\mathcal{F}^{2}\!+\!\mathcal{O}\!\left(\mathcal{F}^{\frac{9}{4}}\right).     
\end{eqnarray}
In the weak-field limit, $\mathcal{F}\rightarrow\mathcal{F}_{0}$, and
$\ln\!\!\left(\!\frac{4\beta_{\ast}^{2}}{  \sigma_{\!\ast}^{2} s^{2} \mathcal{F} }\!\right)\rightarrow0$.
Thus, in this limit Eq. (\ref{Lder}) becomes
%%%%%%%%%%%%%%%%%%%%%%%%%%%%%%%%%%%%%%%%
\begin{equation}\label{Lminder} 
\mathcal{L}\!=\!\frac{\mathcal{F}}{8}-s^{^{\frac{1}{2}}}\!\!\!\left( 1 \!+\! \frac{37}{210\sigma_{\!\ast}} \!+\! \frac{2}{525\sigma_{\!\ast}}\!\right)\!\mathcal{F}^{^{\frac{5}{4}}} - \frac{ \sigma_{\!\ast} s \mathcal{F}^{^{\frac{3}{2}}}}{16}
+\mathcal{O}(\mathcal{F}^{^{\frac{7}{4}}}).
\end{equation}
%%%%%%%%%%%%%%%%%%%%%%%%%%%%%%% 
Hence, under the assumption of the minimum value for $\mathcal{F}_0$, the NLED model (\ref{NLEDt}) reduces in the weak-field limit to that of 
Maxwell's electrodynamics: $\mathcal{L} \!\rightarrow\!\frac{1}{8}\mathcal{F}$, $\mathcal{L}_{\mathcal{F}} \!\rightarrow\!\frac{1}{8}$.
% 

%%%%%%%%%%%%%%%%%%%%%%%%%%%%%%%%%%%%%%%%%%%%%%%%%%%%%%%%%%%%%%%%%%%%%%%%%%%%%%%%%%%%%%%%%%%%%%%%%%%%%%%%%%%%%%%%%%%%
% 
%%%%%%%%%%%%%%%%%%%%%%%%%%%%%%%%%%%
Let us move now to the new exact solution for the model defined by 
the equations given above. 
The ESTGB-NLED model we have presented admits a magnetic ESTGB-NLED exact solution for the case in which its parameters satisfy the relations
\begin{equation}
\sigma = \sigma_{\!\ast} = \frac{q}{m}, \quad\quad\quad \ell=s=q, \quad\quad\quad \beta = \beta_{\ast}.
\end{equation}
For such a solution,  the line element
is given by 
%%%%%%%%%%%%%%%%%%%%%%%%%%%%%%%%%%%%
\begin{equation}\label{NewBH}
ds^{2} =  - \left( 1 - \frac{2m}{r} - \frac{ q^{3} }{r^{3}} \right)dt^{2} + \left( 1 - \frac{2m}{r} - \frac{q^{3}}{r^{3}}  \right)^{\!-1}dr^{2}+ r^{2}(d\theta^{2}  + \sin^{2}\theta d\varphi^{2}),
\end{equation}
and
the electromagnetic invariant is determined by Eq.(\ref{magnetica}).
Finally, the scalar field is simply
\footnote{In order to check that the expressions presented here are a solution of the EOM, it is convenient to have the dependence of the relevant functions with $r$, which is shown in the Appendix.}
%%%%%%%%%%%%%%%%%%%%%%%%%%%%%%%
\begin{equation}\label{scalarex}
\phi(r) = \frac{q}{r}. 
\end{equation}
It follows that the new black hole solution presented here
has scalar hair, which is of the secondary kind, since it is not accompanied by any new quantity
that characterizes the black hole. This is due to the relation between the magnetic and the scalar charge. 

Comparing the metric given in Eq. (\ref{NewBH}) with Eq. (\ref{SSS_ansatz}),  it follows that $\mathscr{M}\!(r) = m + \frac{q^{3}}{2r^{2}}$ and $\delta(r)=0$. Thus, the corresponding ADM mass will be $\mathscr{M}_{_{\!\!A\!D\!M}} = \lim\limits_{r\rightarrow \infty}\mathscr{M}\!(r) = m$. From Eqs.(\ref{scalar})
and
\eqref{scalarex}
it follows that the parameter $q$ corresponds to the scalar charge.
%%%%%%%%%%%%%%%%%%%%%%%%%%%%%%%%%%%%

Evaluating the curvature invariants for this metric, we get 
%%%%%%%%%%%%%%%%%%%%%%%%%%%%%%%%%%%%
\begin{equation}\label{invariants}
R = \frac{2q^{3}}{r^{5}}, \quad R_{\alpha\beta}R^{\alpha\beta} = \frac{26q^{6}}{r^{10}}, \quad R_{\alpha\beta\sigma\nu}R^{\alpha\beta\sigma\nu} = \frac{4(3m^{2}r^{4} + 20mq^{3}r^{2} + 46q^{6})}{r^{10}}.   
\end{equation} 
Hence, analogously to the vacuum (or electrovacuum) AF-SSS black hole solutions in General Relativity, the origin $r=0$ is a physical singularity, and the solution here presented is regular everywhere else. 
%
%%%%%%%%%%%%%%%%%%%%%%%%%%%%%%%%%%%%%%%%%%%%%%%%%%%%%%%%%%%%%%%%%%%%%%%%%%%%%%%%%%%%%%%%%%%%%%%%%%%%%%%%%%%%%%%%%%%
We shall show below that the
NLED satisfies the weak energy condition (WEC), which is defined as follows.
From Eqs. (\ref{E_nled}) and (\ref{diagonalEab}), we obtain
\begin{equation}
 \rho^{^{\!(\!N \! L \! E \! D\!)}} = \frac{\mathcal{L}}{4\pi},\quad\quad\quad\quad P_{r}^{^{\!(\!N \! L \! E \! D\!)}} = -\frac{\mathcal{L}}{4\pi}, \quad\quad\quad\quad P_{\theta}^{^{\!(\!N \! L \! E \! D\!)}} = P_{\varphi}^{^{\!(\!N \! L \! E \! D\!)}} =  \frac{2\mathcal{F}\mathcal{L}_{\mathcal{F}} - \mathcal{L}}{4\pi}.    
\end{equation}
Hence, according to (\ref{WEC}), the NLED energy-momentum tensor (\ref{E_nled}) satisfies the WEC if   
%%%%%%%%%%%%%%%%%%%%%%%%%%%%%%%%%%%%%%%%%%%%%%%%%%%%55
\begin{equation}\label{WECNLED}
\rho^{^{\!(\!N \! L \! E \! D\!)}} = \frac{\mathcal{L}}{4\pi} \geq0, \quad \rho^{^{\!(\!N \! L \! E \! D\!)}} + P^{^{\!(\!N \! L \! E \! D\!)}}_{r} = 0, \quad \rho^{^{\!(\!N \! L \! E \! D\!)}} + P^{^{\!(\!N \! L \! E \! D\!)}}_{a}=\frac{\mathcal{F}\mathcal{L}_{\mathcal{F}} }{2\pi} \geq0, \quad  a = \{ \theta, \varphi \}.  
\end{equation}
Since the invariant $\mathcal{F}$ is positive definite, see Eq.(\ref{magnetica}), then the WEC is holds if $\mathcal{L}\geq0,$ $\mathcal{L}_{\mathcal{F}}\geq0.$ 

It will also be required that the 
effective energy-momentum tensor satisfies the WEC\footnote{See 
\cite{Garcia2010} for the case of modified GB gravity.}.
The components of this tensor in the case of the solution presented here are given by  
%%%
\begin{equation}\label{effectTab}
\left( 8\pi \mathscr{E}_{\nu}{}^{\mu} \right) = {\rm diag}\!\left( \frac{2q^{3}}{r^{5}}, \frac{2q^{3}}{r^{5}}, -\frac{3q^{3}}{r^{5}}, -\frac{3q^{3}}{r^{5}}\right)    .
\end{equation}
Then, the comparison with Eq. (\ref{diagonalEab}) yields $$\rho^{^{\!(\!e\!f\!f\!)}} = -\frac{2q^{3}}{8\pi r^{5}},\;\;\;\;\;\;\; P_{r}^{^{\!(\!e\!f\!f\!)}}=\frac{2q^{3}}{8\pi r^{5}},\;\;\;\;\;\;\; P_{\theta}^{^{\!(\!e\!f\!f\!)}}=P_{\varphi}^{^{\!(\!e\!f\!f\!)}}= -\frac{3q^{3}}{8\pi r^{5}}.
$$
It follows that
$$ \rho^{^{\!(\!e\!f\!f\!)}} + P_{r}^{^{\!(\!e\!f\!f\!)}}=0,\;\;\;\;\;\;\;
{\rm and}\;\;\;\;\;\;\; \rho^{^{\!(\!e\!f\!f\!)}} + P_{\theta}^{^{\!(\!e\!f\!f\!)}}= \rho^{^{\!(\!e\!f\!f\!)}} + P_{\varphi}^{^{\!(\!e\!f\!f\!)}} = -\frac{5q^{3}}{8\pi r^{5}}.$$ 
Therefore, we conclude that if $q\leq0$, the effective energy momentum tensor satisfies the WEC. 
%%%%%%%%%%%%%%%%%%%%%%%%%%%%%%%%%%%%%%%%%%%%%%%%%%%%%%%%%%%%%%%%%%%%%%%%%%%%%%%%%%%%%%%%%%%%%%%%%%%%%
This is in contrast  with the ESTGB Maxwell, and ESTGB power-Maxwell, traversable wormhole solutions presented in \cite{Exact_wH_ED,Exact_wH_NLED}, for which the corresponding effective energy-momentum tensors violate the WEC  
in some regions of the spacetime.
%%%%%%%%%%%%%%%%%%%%%%%%%%%%%%%%%%%%%%%%%%%%%%%%%%%%%%%%%%%%%%%%%%%%%%%%%%%%%%%%%%%%%%%%%%%%%%%%%%%%%%%%%%%%%%%%%%%
%%%%%%%%%%%%%%%%%%%%%%%%%%%%%%%%%%%%

Let us present next the properties of the solution and of the model according to the different possible choices of parameters in them. 

%%%%%%%%%%%%%%%%%%%%%%%%%%%%%%%%%%%%%%%%%%%%%%%%%%%%%%%%%%%%%%

\subsection{ Horizons and black hole interpretations}

We begin by showing  that the metric given in Eq. (\ref{NewBH}) admits several AF-SSS-BH interpretations,
depending on  the range of the values taken by its parameters. The horizons are determined by the roots of the function
$g_{tt}(r)=-(r^{3}-2mr^{2}-q^{3} )/r^{3}$.
In the general case, $g_{tt}$ can have three different roots which  will be denoted by $x_{_{0}}$, $x_{_{1}}$, $x_{_{2}}$. However, a given root $r=x_{i}$ will correspond 
to the radius of the of a horizon only if $x_{i}\in\mathbb{R}^{+}$. The root $x_{_{0}}$ is given by  
%%%%%%%%%%%%%%%%%%%%%%%%%%%%%%%%%%%%%%%%%%%%%%%%%%%%%%%%%%%%%%
\begin{equation}\label{xcero}
x_{_{0}} = \frac{2m}{3} + \frac{(108q^{3} + 64m^{3} + 12\sqrt{ 96m^{3}q^{3} + 81q^{6} })^{\frac{1}{3}}}{6} + \frac{ 8m^{2} }{3(108q^{3} + 64m^{3} + 12\sqrt{96m^{3}q^{3} + 81q^{6}})^{\frac{1}{3}}}.
\end{equation}
$x_{0}$ can be used to
write the metric component $g_{tt}$ as   $g_{tt} = -( r - x_{_{0}} )\!\!\left[r^{2} + (x_{_{0}} - 2m)r + (x_{_{0}}^{2} - 2mx_{_{0}}) \right]/r^{3}$, and  $x_{_{1}}$ and $x_{_{2}}$ are now the roots of the polynomial %%%%%%%%%%%%%%%%%%%%%%%%
\begin{equation}\label{polyn}
\mathscr{P}(r) = r^{2} + (x_{_{0}} - 2m)r + (x_{_{0}} - 2m)x_{_{0}},     
\end{equation}
%%%%%%%%%%%%%%%%%%%%%%%
% 
which
are given by
\begin{equation}
\label{r12}
x_{_{1}} = \frac{ 2m - x_{_{0}} + \sqrt{ - 3 x_{_{0}}^{2} + 4mx_{_{0}} + 4m^{2}  } }{2}, \quad\quad\quad\quad\quad  x_{_{2}} = \frac{ 2m - x_{_{0}} - \sqrt{ - 3 x_{_{0}}^{2} + 4mx_{_{0}} + 4m^{2}  } }{2}.     
\end{equation}
Several cases must be considered, according to the sign of the parameters involved:

\subsubsection{Case $\{ m>0$, $q>0 \}$}
\begin{figure}
\centering
\epsfig{file=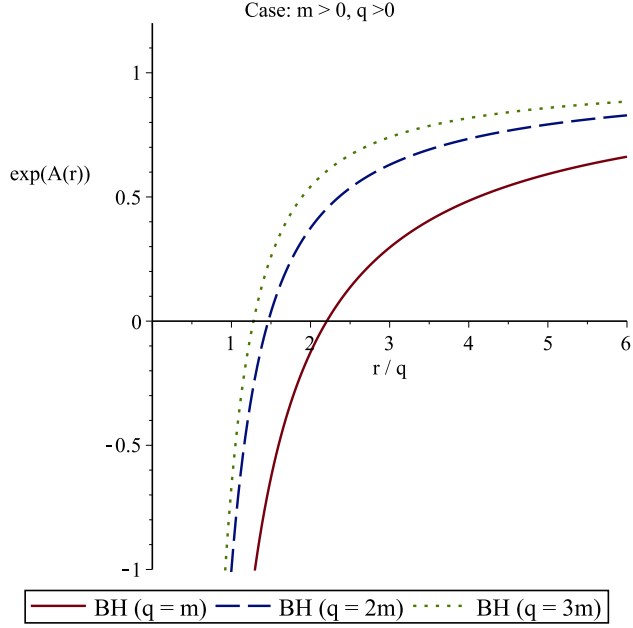, scale=0.44}
\caption{\label{Schw} Behavior of  $-g_{tt}$ for $m>0$ and $q>0$, with different values of 
$q/m$. In all cases, there is a single horizon.
}
\end{figure}

For this  setting of parameters, it follows from Eq.
\eqref{xcero} that $x_0$ is real. Hence, it corresponds to the radius of a horizon.
Also, 
the following inequality is valid:
\begin{equation}
x_{_{0}} = \frac{2m}{3} + \frac{(108q^{3} + 64m^{3} + 12\sqrt{ 96m^{3}q^{3} + 81q^{6} })^{\frac{1}{3}}}{6} + \frac{ 8m^{2} }{3(108q^{3} + 64m^{3} + 12\sqrt{96m^{3}q^{3} + 81q^{6}})^{\frac{1}{3}}} \quad \geq \quad 2m   
\end{equation}
%%%%%%%%%%%%%%%%%%%%%%%%%%%%%%%%%%%%%%%%%%%%%%%%%%%%
%
Given that $x_{_{0}} \geq 2m>0$ the roots of the polynomial $\mathscr{P}(r)$, given by Eq.\eqref{r12} do not correspond to positive real numbers. 
Thus for this case the metric 
in Eq. (\ref{NewBH}) admits a AF-SSS-BH interpretation, with structure similar to the Schwarzschild metric, \emph{i.e.}, positive ADM mass, only one event horizon at the surface $r=x_{_{0}}$, and 
satisfying 
$g_{tt}(r)\rightarrow\infty$ as $r\rightarrow0$. 
The plot of  $-g_{tt}(r)$ in terms of the dimensionless coordinate $r/q$ 
is displayed in 
Fig. \ref{Schw}
for 
different values of $q/m$.

\subsubsection{Case  $\{ m>0$, $q<0 \}$}

\begin{figure}
\centering
\epsfig{file=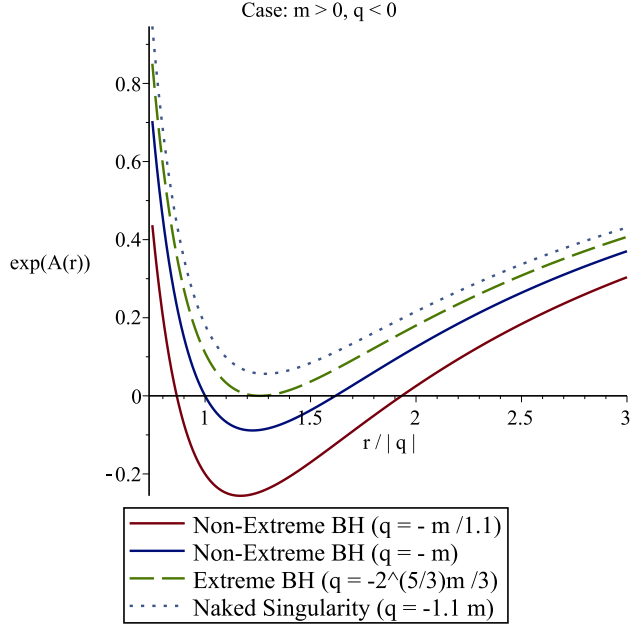, scale=0.44}
\caption{\label{RN} Behavior of  $-g_{tt}$ for $m>0$ and $q<0$,  with different values of $m/q$.
}
\end{figure}

This case encompasses three sub-cases characterized by 
whether 
$96m^{3}q^{3} + 81q^{6}$ is negative, null, or positive. 
For the sub-case $96m^{3}q^{3} + 81q^{6}<0$, we can define 
%%%%%%%%%%%%%%%%%%%
\begin{eqnarray}
&&z_{_{0}} = 108q^{3} + 64m^{3} + 12i\sqrt{ - 96m^{3}q^{3} - 81q^{6} } = \rho_{_{0}}e^{i\theta_{_{0}}} \in \mathbb{C}, \hskip.6cm \textup{with} \hskip.4cm \rho_{_{0}} = 64m^{3}, \nonumber \\ 
&&  \theta_{_{0}} \!=\! \tan^{^{-1}}\!\!\left( \frac{3\sqrt{ -96q^{3}m^{3} - 81q^{6} }}{|27q^{3} + 16m^{3}|} \right) \hskip.3cm \textup{if} \hskip.3cm \mathcal{R}e(z_{_{0}})>0, \hskip.3cm \theta_{_{0}} \!=\! \pi \!-\! \tan^{^{-1}}\!\!\left( \frac{3\sqrt{ -96q^{3}m^{3} - 81q^{6} }}{|27q^{3} + 16m^{3}|} \right) \hskip.3cm \textup{if} \hskip.3cm \mathcal{R}e(z_{_{0}})<0.
\end{eqnarray}
Hence,  $0<\theta_{_{0}}<\pi$, and $x_{_{0}}$ can be written as
\begin{equation}\label{outerEH}
x_{_{0}} =  \frac{2m}{3} + \frac{1}{6} \rho^{ \frac{1}{3}}_{_{0}} e^{ \frac{1}{3} i \theta_{_{0}}  }  + \frac{ 8m^{2} }{3}\rho^{ -\frac{1}{3}}_{_{0}} e^{ - \frac{1}{3} i \theta_{_{0}}  }  = \frac{2m}{3}\!\!\left[ 1 + 2\cos\left( \frac{\theta_{_{0}}}{3}\right) \right] .
\end{equation}
leading to
$\frac{4m}{3} < x_{_{0}} < 2m$.
Due to this restriction on $x_0$,
the polynomial $\mathscr{P}(r)$ has two different real zeros, $x_{_{1}}$ and $x_{_{2}}$, given respectively by,  
\begin{equation}
x_{_{1}} = \frac{ 2m - x_{_{0}} + \sqrt{ - 3 x_{_{0}}^{2} + 4mx_{_{0}} + 4m^{2}  } }{2}>0, \quad \quad \quad  x_{_{2}} = \frac{ 2m - x_{_{0}} - \sqrt{ - 3 x_{_{0}}^{2} + 4mx_{_{0}} + 4m^{2}  } }{2} < 0     
\end{equation}
with $0 < x_{_{1}} < \frac{4m}{3}$. It follows that for this sub-case, in general way, the solution supports two horizons located at $r_{h}=x_{_{0}}$ (outer event horizon) and $r_{_{1}}=x_{_{1}}$ (inner horizon). Hence, this sub-case leads a non-extreme BH. 
On another hand, the sub-case $96m^{3}q^{3} + 81q^{6}=0$ (which is equivalent to $\theta_{_{0}}=3\pi$),  leads to $x_{_{0}}=-\frac{2}{3}m$, $r_{_{1}} = r_{_{2}} = x_{_{1}} = x_{_{2}} = \frac{4m}{3}$, which corresponds to an extreme black hole with event horizon given by $r_{h} = \frac{4m}{3}$. Finally,  the sub-case $96m^{3}q^{3} + 81q^{6}>0$ yields $x_{_{0}}$, $x_{_{1}}$, $x_{_{2}}$ $\not \in \mathbb{R}^{+}$, and then only for pairs $(m,q)$ that satisfy this condition the solution does not present an event horizon, leading to a naked singularity at $r=0$. Summarizing, when $\{ m>0$, $q<0 \}$, our solution has characteristics similar to those of the Reissner-Nordstr\"om metric. Fig. \ref{RN} shows the plot of $-g_{tt}(r)$ for a fixed values of $m/q$, 
corresponding to solutions a non-extreme BH, an extreme BH, and a naked singularity.

\subsubsection{Case  $\{ m<0$, $q>0 \}$}

Within this case, we consider the sub-case $96m^{3}q^{3} + 81q^{6}\geq0$. For this setting of parameters, the quantity $x_{_{0}}$ given by Eq.(\ref{xcero}) is such that  $x_{_{0}}\geq\frac{2|m|}{3}$. This implies that the roots of the polynomial
%%%%%%%%%%%%%%%%%%%%%%%%%%%%%%%%%%%%%%%
$\mathscr{P}(r) = r^{2} + (x_{_{0}} - 2m)r + (x_{_{0}}^{2} - 2mx_{_{0}})$, are not positive real numbers. Whereas, if $96m^{3}q^{3} + 81q^{6}\leq0$, 
we can define
%%%%%%%%%%%%%%%%%%%%%%%%%%%%%%%%%%%%%%%
\begin{eqnarray}
&&z_{_{0}}\!=\!108q^{3} + 64m^{3} + 12i\sqrt{ - 96 q^{3} m^{3} - 81q^{6} }\!=\!\rho_{_{0}}e^{i\theta_{_{0}}}, \hskip.6cm \textup{with} \hskip.4cm \rho_{_{0}}\!=\!-64m^{3},\\
&& \theta_{_{0}}\!=\!\tan^{\!^{-1}}\!\!\left( \frac{3\sqrt{ - 96 q^{3} m^{3} - 81q^{6} }}{ |27q^{3} + 16m^{3}| }\right)\hskip.3cm \textup{if} \hskip.3cm \mathcal{R}e(z_{_{0}})>0, \hskip.3cm    \theta_{_{0}}\!=\!\pi\!-\! \tan^{\!^{-1}}\!\!\left( \frac{3\sqrt{ - 96 q^{3} m^{3} - 81q^{6} }}{ |27q^{3} + 16m^{3}| }\right)\hskip.3cm \textup{if} \hskip.3cm \mathcal{R}e(z_{_{0}})<0,  
\end{eqnarray}
leading to $0<\theta_{_{0}}<\pi$. Hence, for this sub-case $x_{_{0}}$ 
is given by
\begin{equation}
x_{_{0}} =  \frac{2m}{3} + \frac{1}{6} \rho^{ \frac{1}{3}}_{_{0}} e^{ \frac{1}{3} i \theta_{_{0}}  }  + \frac{ 8m^{2} }{3}\rho^{ -\frac{1}{3}}_{_{0}} e^{ - \frac{1}{3} i \theta_{_{0}}  }  = - \frac{2m}{3}\left( 2\cos\left( \frac{\theta}{3}\right) - 1 \right),
\end{equation}
which leads to 
$x_{_{0}}<\frac{2|m|}{3}$.
Thus, since $x_{_{0}}>0$ and $m<0$, the roots of the polynomial $\mathscr{P}(r)$  do not correspond to positive real numbers.

\begin{figure}
\centering
\epsfig{file=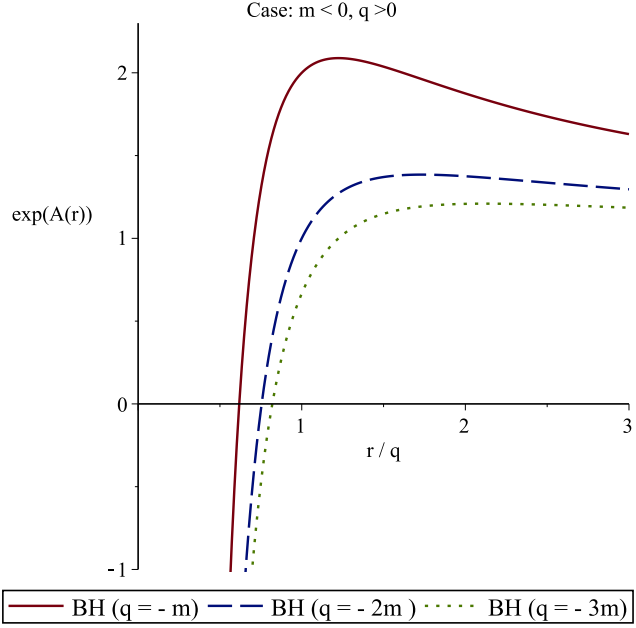, scale=0.44}
\caption{\label{SCH_Mneg} Behavior of  $-g_{tt}$ for $m<0$ and $q>0$, with different values of $q/m$.
}
\end{figure}

Therefore, for the case $\{ m<0$, $q>0 \}$, the solution (\ref{NewBH}) can be interpreted as an AF-SSS-BH with negative ADM mass \footnote{Black holes with negative mass have been considered for instance in 
\cite{Mann1997,Martinez2005}.}
and only one horizon at $r=x_{_{0}}$. Plots of $-g_{tt}(r)$
for this case are presented in Fig. \ref{SCH_Mneg}, for different values of $q/m$.

\subsubsection{Case  $\{ m<0$, $q<0 \}$}

For this case, all the roots of $g_{tt}(r)$ 
do not belong to 
$\mathbb{R}^{+}$. Then for this set of parameters the metric (\ref{NewBH}) represents a naked singularity .
%
%%%%%%%%%%%%%%%%%%%%%%%%%%%%%%%%%%%%%%%%%%%%%%%%%%%%%%%%%%%%%%%%%%%%%%%%%%%%%%%%%%%%%%%%%%%%%%%%%%%%%%%%%%%%%%%%%%%%%
%
%

\subsubsection{Restrictions imposed by the weak energy condition} 

Let us see now the restrictions that follow from taking into account that 
both the energy-momentum tensor associated to the NLED Lagrangian and the effective energy-momentum tensor have to fulfill the conditions dictated by the WEC. First, notice that following the discussion after Eq. (\ref{effectTab}), only solutions with $q<0$ are
acceptable. Thus, 
from now on we shall restrict the analysis to the 
case ${m>0, q<0}$, since the rest of the cases lead to an effective energy-momentum tensors which does not satisfy the WEC. Consequently, the notation $Q=-q$ will be used from now on, with $Q>0$.

Regarding the WEC and the NLED energy-momentum tensor, we show next that the corresponding local NLED energy density is positive definite outside the black hole event horizon.
In order to describe the behavior of 
$\mathcal{L}$ and $\mathcal{L}_{\mathcal{F}}$, we shall introduce the variable $\tilde{\mathcal{F}}$ defined by the rescaling, $\tilde{\mathcal{F}} = q^{2} \mathcal{F}$. Using Eq. (\ref{magnetica}) and $\tilde{\mathcal{F}}$ we can rewrite $\mathcal{L}$ given in Eq. (\ref{LNewrQ}) as a function
of $\tilde{\mathcal{F}}$. The result is
%%%%%%%%%%%%%%%%
\begin{eqnarray}
Q^{2}\mathcal{L}&=&\frac{\tilde{\mathcal{F}}}{8}\!+\!\frac{8m^{3}}{105Q^{3}} \tilde{\mathcal{F}}^{\frac{1}{4}}\!+\!\frac{4m^{2}}{315Q^{2}}\tilde{\mathcal{F}}^{\frac{3}{4}}\!-\!\left(\!\frac{37m}{210Q}\!-\!1\!\right)\!\tilde{\mathcal{F}}^{\frac{5}{4}} \!+\!\frac{5}{84}\!\tilde{\mathcal{F}}^{\frac{7}{4}}\!-\!\frac{ 2^{\!^{\frac{3}{2}}} m^{ \!^{\frac{7}{2}} } }{ 105 Q^{\frac{7}{2}}  }\!\ln\!\!\!\left[\!\!\left(\!\!\frac{ \frac{\tilde{\mathcal{F}}^{^{\frac{1}{4}}} }{ 2^{^{\! \frac{1}{2}}} }\!+\!\frac{ m^{^{\!\frac{1}{2}}} }{ Q^{^{\!\frac{1}{2}}} }  }{ \frac{\tilde{\mathcal{F}}^{^{\frac{1}{4}}}}{ 2^{^{\! \frac{1}{2}}} }\!-\! \frac{ m^{^{\!\frac{1}{2}}} }{ Q^{^{\!\frac{1}{2}}} }  }\!\!\right)^{\!\!\!2}\right] \nonumber \\
&&\hskip5.2cm -\frac{1}{2}\!\left(\!\frac{\tilde{\mathcal{F}}}{8}\!-\!\frac{3m}{10Q}\tilde{\mathcal{F}}^{^{\frac{5}{4}}}\!-\!\frac{3Q}{16m}\tilde{\mathcal{F}}^{^{\frac{3}{2}}}\!+\!\frac{4}{7}\tilde{\mathcal{F}}^{^{\frac{7}{4}}}\!-\!\frac{5Q}{24m}\tilde{\mathcal{F}}^{^{\frac{9}{4}}}\!\right)\!\ln\!\!\left[\!\!\left(\!\frac{2m\beta}{Q\tilde{\mathcal{F}}^{^{\frac{1}{2}}}}\!-\!\beta\!\right)^{\!\!2}\right]\!\!,\label{LmQ} %
\end{eqnarray}
The values of $r$ outside the black hole, given by
$r\in (r_h, \infty)$, with $r_h$ the outer event horizon, given by $r_{h}=x_{_{0}}$, see Eq. 
(\ref{outerEH}),
%%%%%%%%%%%%%%%%%%%%%%%%%%%%%
or the single event horizon in the extremal case, give by $r_{h}=\frac{4m}{3}$, delimitate the
domain of $q^2{\cal L}$, which is given by
$\tilde{D} = \left[\tilde{\mathcal{F}}_{_{0}}, \tilde{\mathcal{F}}_{_{h}}\right]$, where $\tilde{\mathcal{F}}_{_{0}} = q^{2}\mathcal{F}_{_{0}} = \frac{ 4\beta^{2} m^{2} }{ q^{2} }$ is the minimum value of the rescaled variable, and $\tilde{\mathcal{F}}_{_{h}} = q^{2}\mathcal{F}_{_{h}} = \frac{q^{4}}{ r^{4}_{_{h}} }$ its value at the event horizon. 

\begin{figure}
 \centering
{
   \label{f:a}
    \includegraphics[width=0.35\textwidth]{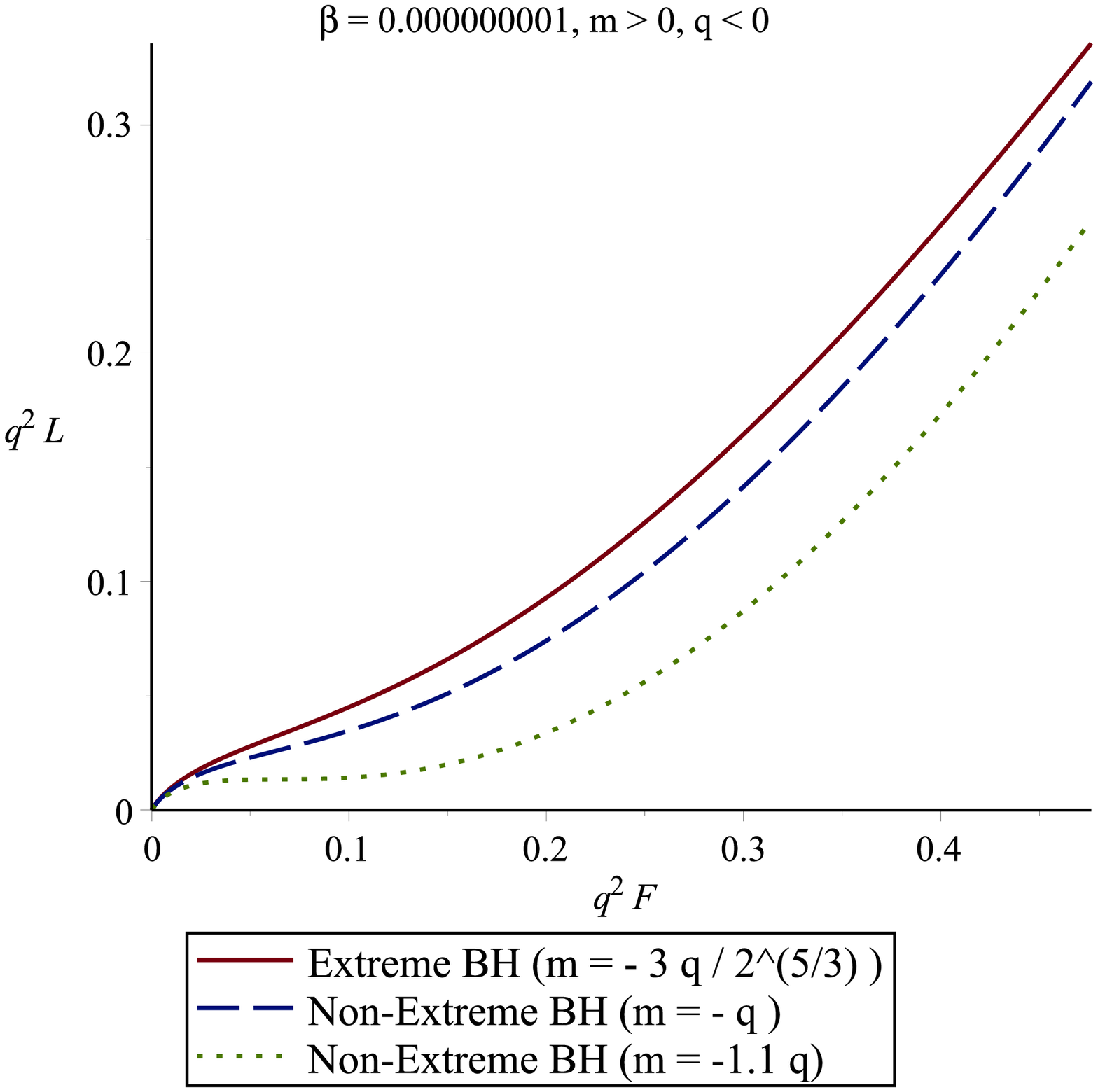}}
{
   \label{f:b}
    \includegraphics[width=0.35\textwidth]{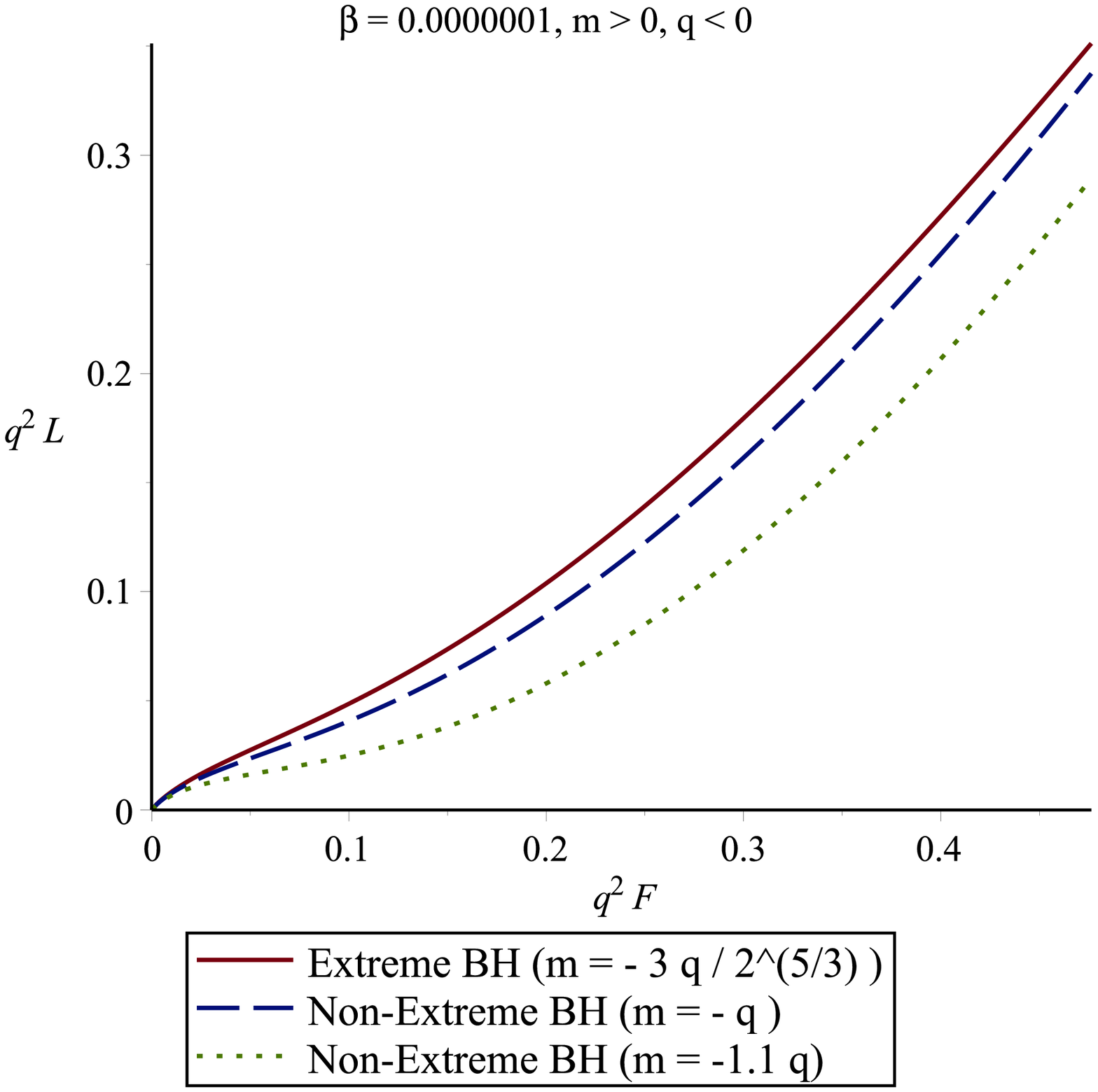}}
  {
   \label{f:c}
    \includegraphics[width=0.35\textwidth]{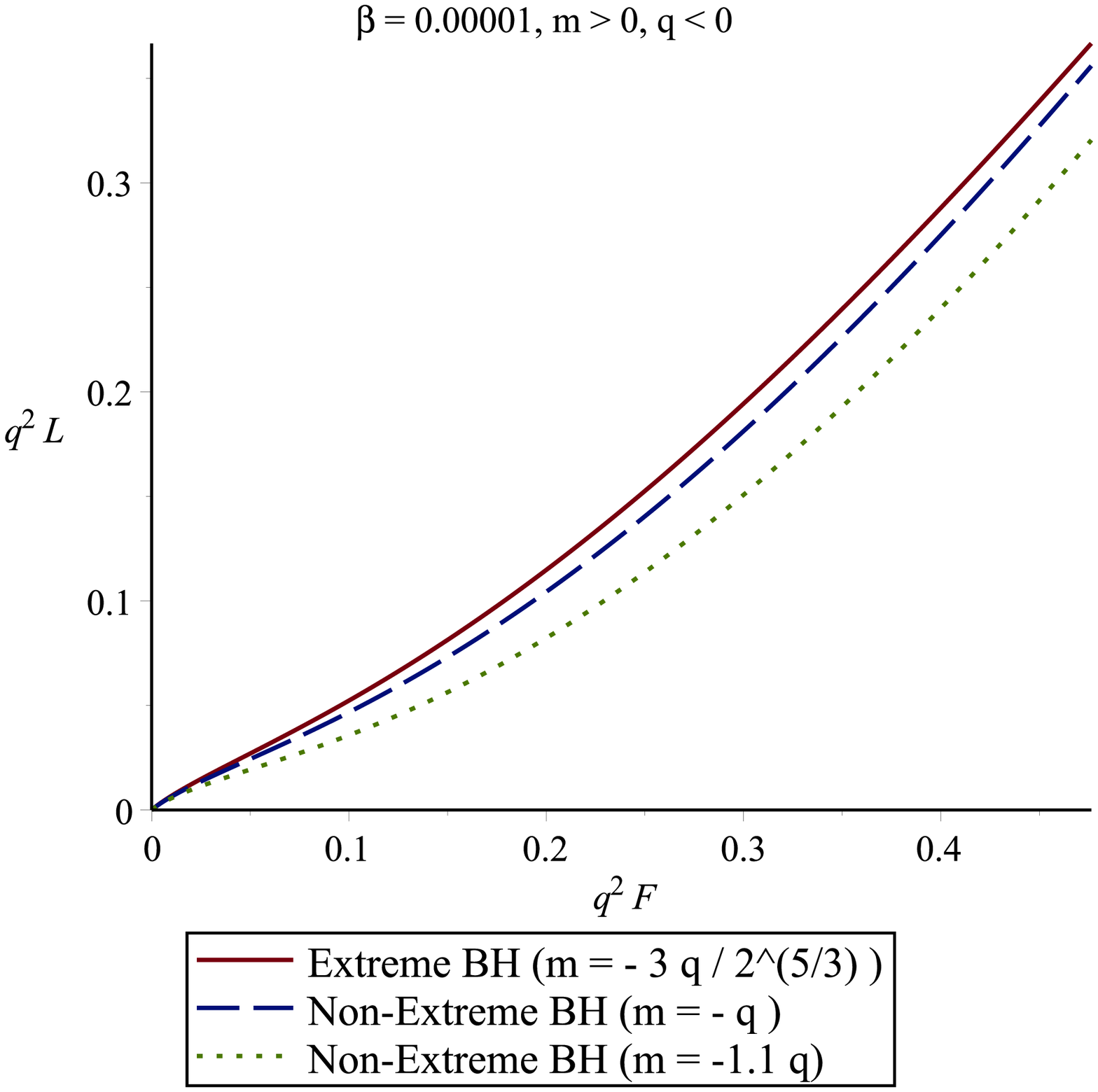}}
 \caption{ Behavior of $q^{2}\mathcal{L}$ as a function of $q^{2}\mathcal{F}$ for different values of $m/q$.}
 \label{LWEC}
\end{figure}

\begin{figure}
 \centering
{
   \label{f:a}
    \includegraphics[width=0.35\textwidth]{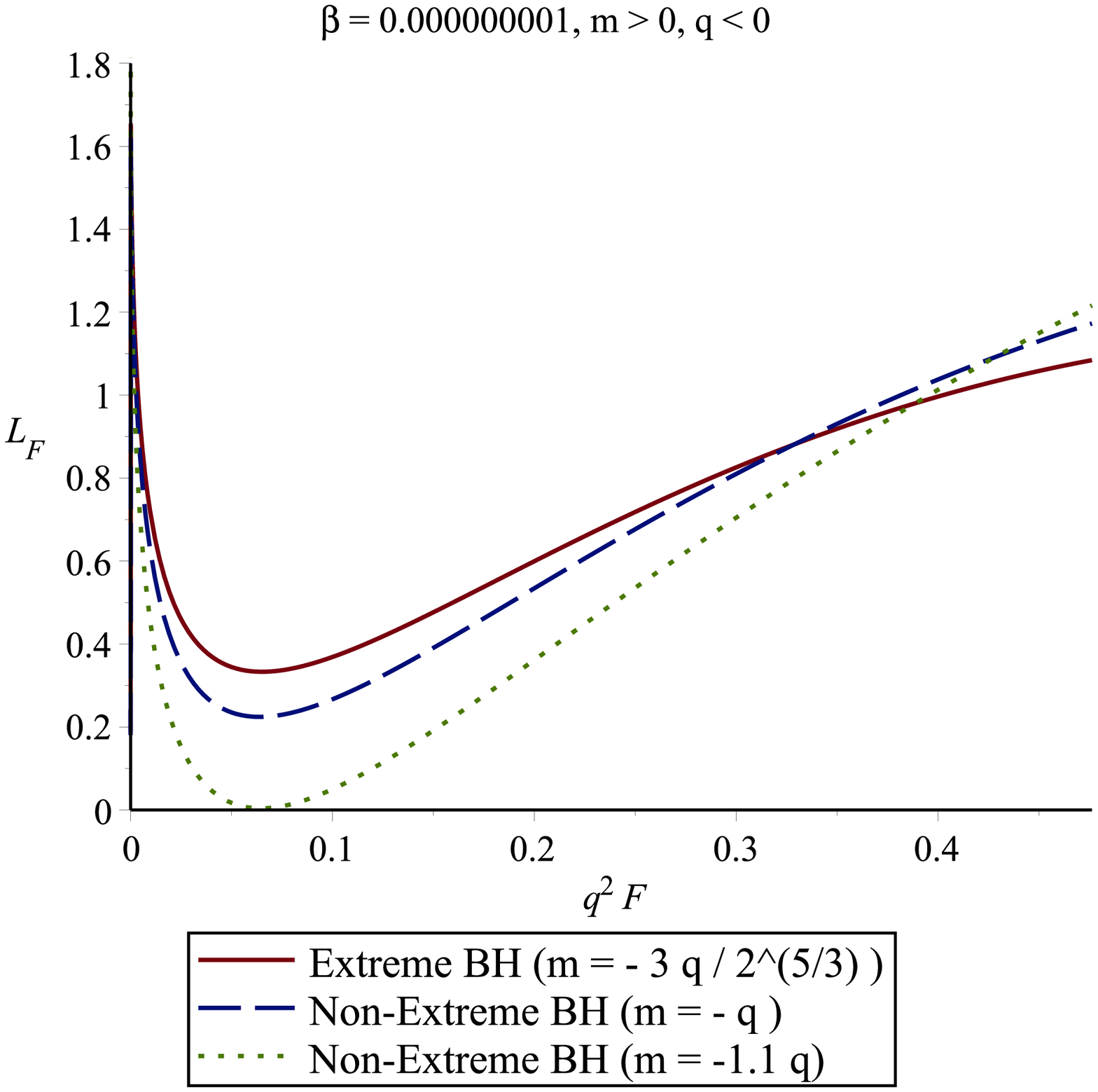}}
{
   \label{f:b}
    \includegraphics[width=0.35\textwidth]{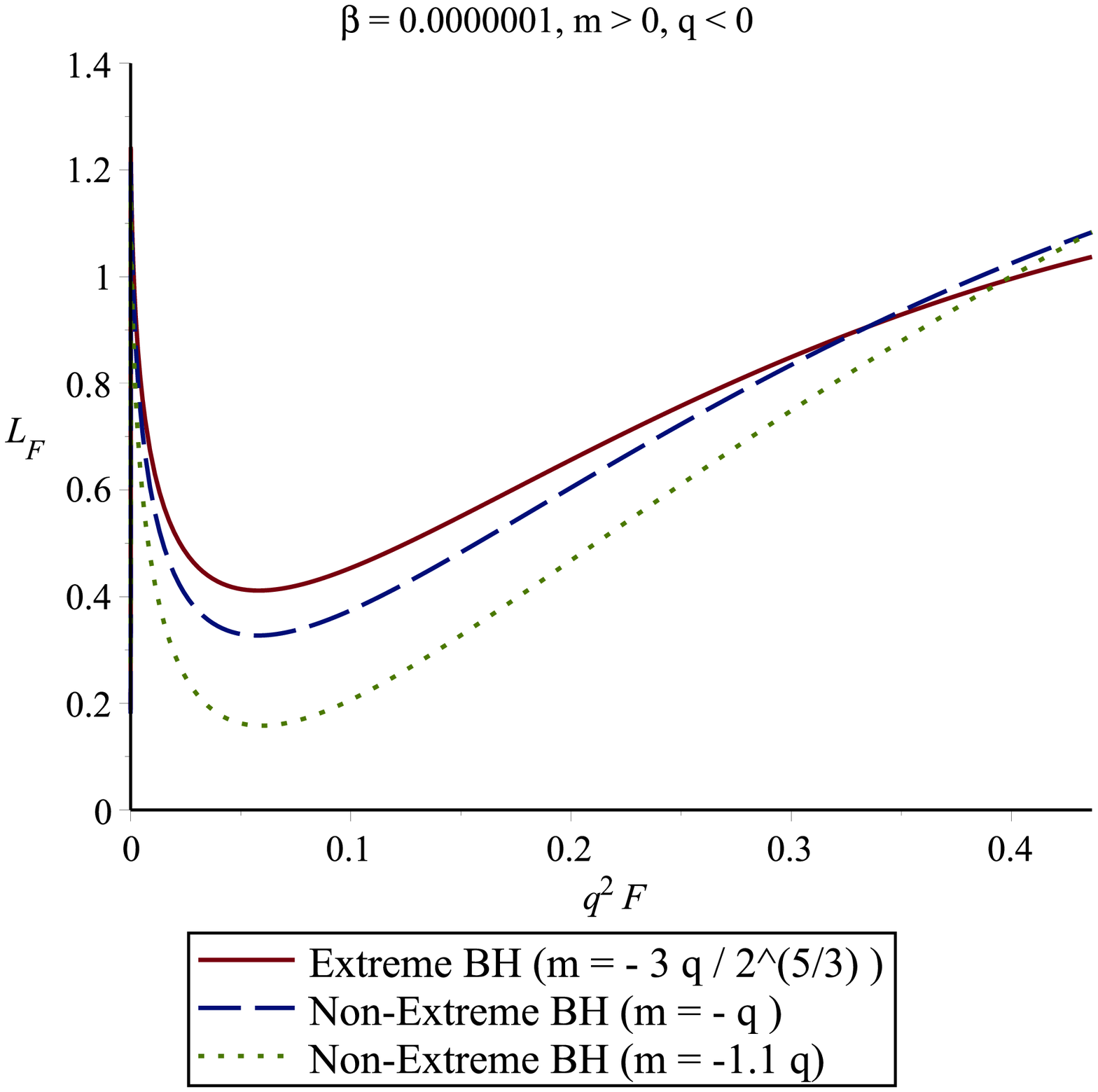}}
  {
   \label{f:c}
    \includegraphics[width=0.35\textwidth]{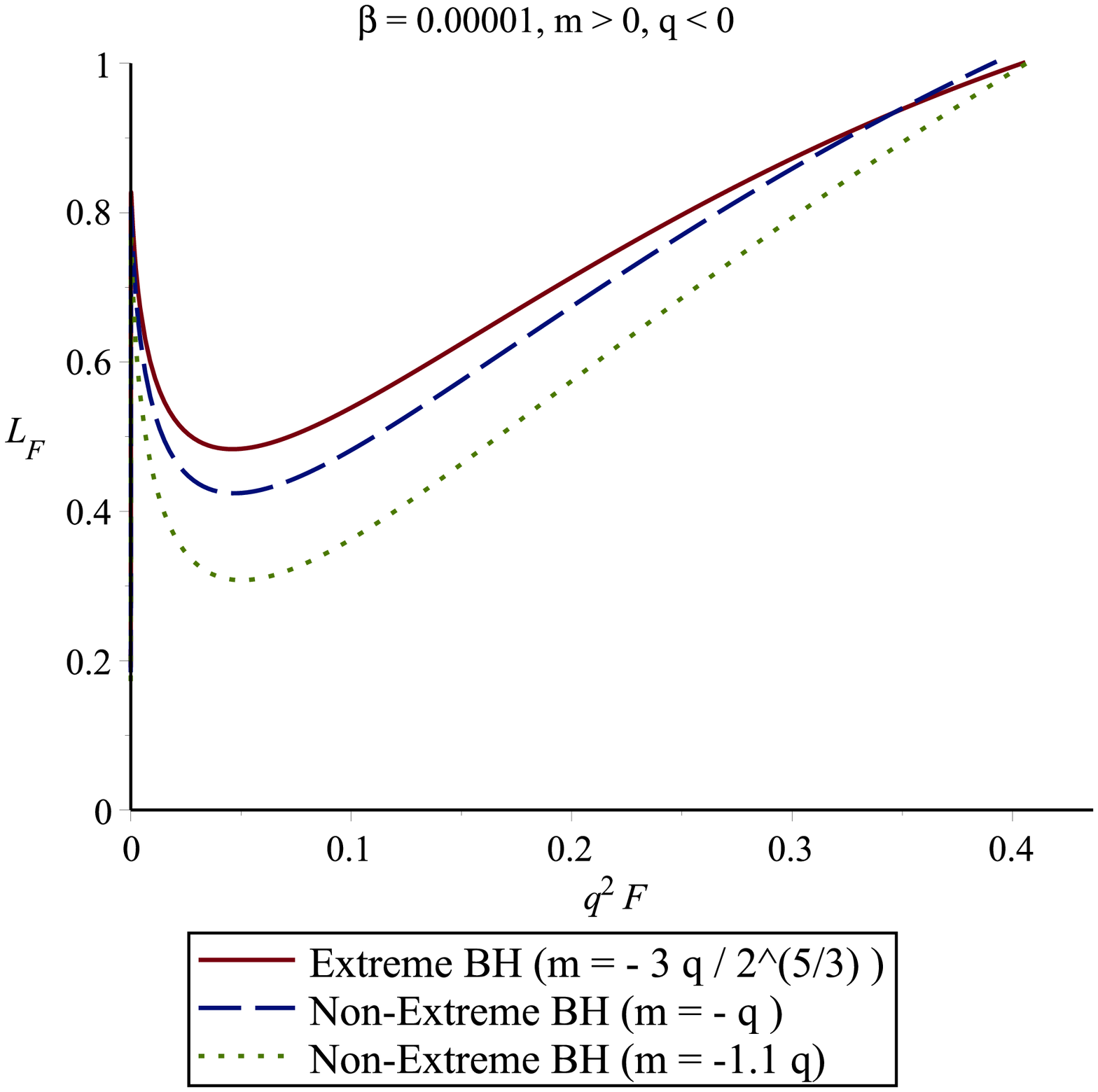}}
 \caption{ Behavior of $\mathcal{L}_{\mathcal{F}}$  as a function of $q^{2}\mathcal{F}$ for different values of $m/q$. }
 \label{LFWEC}
\end{figure}

As examples of the general behavior, 
we shall use the values of the parameters in Fig. \ref{RN}, which lead to the following families of solutions: 
%%%%%%%%%%%%%%%%%%%%%%%%%%%%%%%%%
\begin{enumerate}
    \item  $m=-\frac{3q}{ 2^{^{\frac{5}{3}}} } = \frac{3Q}{2^{^{\frac{5}{3}}}}$ defines a family of extreme BH solutions with $\tilde{\mathcal{F}}$-values 
    outside the BH such that $\tilde{\mathcal{F}}\in\tilde{D}_{_{\!(\!a\!)}}=\left[ \frac{9\beta^{2}_{_{\!(\!a\!)}}}{2^{^{\frac{4}{3}}}}, \frac{1}{2^{^{\frac{4}{3}}}} \right]$, with $\beta^{2}_{_{\!(\!a\!)}}<\frac{1}{9}$ (which follows from imposing that 
$    \tilde{\mathcal{F}}_{_{0}}< \tilde{\mathcal{F}}_{_{h}}$).
    
%%%%%%%%%%%%%%%%%%%%%%%%%%%%%%%%%%%%%%%%%%%%%%%%%%%%%%%%%%%%%%%%%%%%%%%%%%%%%%%%%%%%%%%%%%%%%%%%%%%%    
    \item  $m=-q=Q$ defines a family of non-extreme BH solutions  with $\tilde{\mathcal{F}}$-values  
    outside the BH given by $\tilde{\mathcal{F}}\in\tilde{D}_{_{\!(\!b\!)}}=\left[ 4\beta^{2}_{_{\!(\!b\!)}}, \frac{4^{2}}{(1+\sqrt{5})^{4}} \right]$, with $\beta^{2}_{_{\!(\!b\!)}}< \frac{4}{(1+\sqrt{5})^{4}}$. 
%%%%%%%%%%%%%%%%%%%%%%%%%%%%%%%%%%%%%%%%%%%%%%%%%%%%%%%%%%%%%%%%%%%%%%%%%%%%%%%%%%%%%%%%%%%%%%%%%%%%    
    \item  $m=-1.1q=1.1Q$ defines a family of non-extreme BH solutions with $\tilde{\mathcal{F}}$-values  
    outside the BH given by $\tilde{\mathcal{F}}\in\tilde{D}_{_{\!(\!c\!)}}=\left[ 4.84\beta^{2}_{_{\!(\!c\!)}}, 0.07175581593\right]$, with $\beta^{2}_{_{\!(\!c\!)}} < 0.01482558180$. 
\end{enumerate}
We shall choose as examples the values
$\beta^{2}_{_{\!(\!a\!)}}=\beta^{2}_{_{\!(\!b\!)}}=\beta^{2}_{_{\!(\!c\!)}}=\beta = 10^{-9}$, $10^{-7}$, and $10^{-5}$. For such values, the inequalities given above are very-well satisfied, and $\tilde{\mathcal{F}}_0$ is small enough to be below any measurable field. The functions $q^{2}\mathcal{L}$, and $\mathcal{L}_{\mathcal{F}}$,  corresponding to each of the above-given families of solutions are exhibited in Fig. \ref{LWEC} and \ref{LFWEC}, respectively. They show that $\rho^{^{\!(\!N \! L \! E \! D\!)}}_{\!_{_{loc}}}$ is equal or greater than zero outside the black hole.

%%%%%%%%%%%%%%%%%%%%%%%%%%%%%%%%%%%%%%%%%%%%%%%%%%%%%%%%%%%%%%%%%%%%%%%%%%%%%%%%%%%%%%%%%%%%%%%%%%%%%%%%%%%%%%%%

%

\subsection{Behavior of the ESTGB-NLED model }
%%%%%%%%%%%%%%%%%%%%%%%%%
In this subsection, we address two particular reductions of the ESTGB-NLED model defined by the functions given in Eqs. (\ref{fNew}), (\ref{UNew}) and (\ref{NLEDt}), and the behavior of the function $f(\phi)$.
%
%%%%%%%%%%%%%%%%%%%%%%%%%
%%%%%%%%%%%%%%%%%%%%%%%%%
%

\subsubsection{Reduction
to General Relativity}

From the following development in power series around $q\approx0$, 
\begin{equation}\label{arct_q0}
\frac{\sqrt{2}\hskip.05cm q^{2}}{\sqrt{qm}}\tan^{\!\!\!^{-1}}\!\!\!\left(\!\!\frac{\sqrt{2} \hskip.05cm mr }{ q\sqrt{qm}}\!\!\right) = \frac{q^{\frac{3}{2}} \pi}{\sqrt{2m}} - \frac{q^{3}}{mr} + \frac{q^{6}}{6m^{2}r^{3}} - \frac{q^{9}}{20m^{3}r^{5}} + \mathcal{O}\!\left(q^{15}\right) ,
\end{equation}
we see that 
$$
\frac{\sqrt{2}\hskip.05cm q^{2}}{\sqrt{qm}}\tan^{\!\!\!^{-1}}\!\!\!\left(\!\!\frac{\sqrt{2} \hskip.05cm mr }{ q\sqrt{qm}}\!\!\right)\Bigg|_{q=0} = 0.
$$
Hence, taking the limit $q\rightarrow 0$ 
in Eq.(\ref{f_r_New}), we get that 
$\boldsymbol{f}(\phi)=0$ for $q=0$. 
For the potential $\mathscr{U}$, 
we have that
\begin{equation}\label{Uq0}
\lim\limits_{ q \rightarrow q_{_{0}} }\mathscr{U}\!(r) = \frac{ 2^{\!^{\frac{7}{2}}} \pi m^{\!^{\frac{7}{2}}} }{ 105q^{\!^{\frac{11}{2}}}_{_{0}} } - \frac{ 16m^{3} }{ 105q_{_{0}}^{4} r } \!+\!\frac{8m^{2}}{315q_{_{0}}r^{3}}\!-\!\frac{16\sqrt{2}m^{4}}{105q_{_{0}}^{5}\sqrt{ q_{_{0}}m } }\!\tan^{\!^{\!-1}}\!\!\!\left(\!\!\frac{\sqrt{2}mr}{ q_{_{0}}\sqrt{q_{_{0}}m}}\!\!\right)\!\!,     
\end{equation}
where we have defined $q_{_{0}}=0$. Using (\ref{arct_q0}), it follows that   
\begin{equation}\label{arctan_q0}
\frac{ 16\sqrt{2} m^{4} }{105q^{5}_{_{0}}\sqrt{q_{_{0}}m}}\tan^{\!^{\!-1}}\!\!\!\left( \frac{\sqrt{2} \hskip.05cm mr }{ q_{_{0}}\sqrt{q_{_{0}}m}}\right) = \frac{ 2^{\!^{\frac{7}{2}}} \pi m^{\!^{\frac{7}{2}} } }{ 105q^{\frac{11}{2}}_{_{0}} } - \frac{ 16m^{3} }{ 105q^{4}_{_{0}}r } + \frac{8m^{2}}{315q_{_{0}}r^{3}}. 
\end{equation}
Inserting Eq. (\ref{arctan_q0}) in Eq.(\ref{Uq0}), we get $\lim\limits_{ q \rightarrow q_{_{0}} }\mathscr{U}\!(r) = 0$. 
%%%%%%%%%%%%%%%%%%%%%%%%%%%%
Proceeding in the same way with Eq.(\ref{LNewr}),  
\begin{equation}
\lim\limits_{ q \rightarrow q_{_{0}} }2\mathcal{L}(r) \!=\! \frac{16m^{3}}{105q_{_{0}}^{4} r }\!-\!\frac{8m^{2}}{315q_{_{0}} r^{3}}\!+\!\frac{16\sqrt{2}m^{4} }{ 105 q_{_{0}}^{5} \sqrt{q_{_{0}}m} }\!\tan^{\!^{\!-1}}\!\!\!\left(\!\!\frac{\sqrt{2} mr }{ q_{_{0}}\sqrt{q_{_{0}}m} }\!\!\right)\!-\!\frac{ 2^{\!^{\frac{7}{2}}} \pi m^{\!^{\frac{7}{2}}} }{ 105q^{\frac{11}{2}} }.
\end{equation}
Using Eq.(\ref{arctan_q0}) we obtain  $\lim\limits_{ q \rightarrow q_{_{0}} }2\mathcal{L}(r)=0$.
%%%%%%%%%%%%%%%%%%%%%%%%%%%%%%%%%%%%%%%%%%%%%%%
Thus, we can conclude that when $q=0$, the theory presented here reduces to vacuum GR, and Eq.(\ref{NewBH}) becomes to the Schwarzschild metric.\\

%%%%%%%%%%%%%%%%%%%%%%%%%%%%%%%%%%%%%%%%%%%%%%%
%

\subsubsection{Reduction to a black hole without ADM mass}

In the case $m=0$ with $q>0$, the relevant components of the metric given in Eq. (\ref{NewBH})
are $-g_{tt}=1/g_{rr}=1-q^{3}/r^{3}$. Hence, the event horizon is at $r=r_{h}=q$. The ADM mass of such a solution is null, and its {\it magnetic charge}/$\sqrt{2}$ equals its scalar charge, given by $q$. 
%%%%%%%%%%%%%%%%%%%%%%%%%%%%%%%%%%%%%%%%%%%%%%%%
Let us see in detail how the relevant functions that define the theory behave in the zero mass limit.
To achieve this goal, we define a quantity $m_{_{0}}$, and use the power series expansions of the functions $\boldsymbol{f}$, $\mathscr{U}$ and $\mathcal{L}$, for small 
$m_0$, thus obtaining  
%%%%%%%%%%%%%%%%%%%%%%%%%%%%%%%%%%%%%%%%%%%%%%%%
\begin{equation}\label{regularizacion}
\lim\limits_{m\rightarrow m_{_{0}}}\!\!\!\!\boldsymbol{f}\!=\!-\frac{r^{3}}{48q} \!-\! \frac{ q^{2}r \ln(\beta)}{ 32m_{_{0}} },\hskip.2cm \lim\limits_{m\rightarrow m_{_{0}}}\!\!\!\!\mathscr{U}=\!\frac{9q^{5}}{56r^{7}}\!+\! \frac{7q^{8}\ln(\beta)}{48m_{_{0}}r^{9}}, \hskip.2cm \lim\limits_{ m \rightarrow m_{_{0}} }\!\!\!\!\mathcal{L}\!=\!-\frac{q^{2}}{4r^{4}}\!-\!\frac{q^{3}}{r^{5}}\!+\!\frac{ 5q^{5} }{14r^{7}}\!-\!\left(\!\!\frac{3q^{5} }{16m_{_{0}}r^{6}}\!-\!\frac{5q^{8} }{ 24m_{_{0}}r^{9} }\!\!\right)\!\ln(\beta). 
\end{equation}
By direct calculation it can be checked that the model  $\boldsymbol{f}=-\frac{r^{3}}{48q} - \frac{ q^{2}r \ln(\beta)}{ 32m_{_{0}} },$ $\mathscr{U} = \frac{9q^{5}}{56r^{7}} + \frac{7q^{8}\ln(\beta)}{48m_{_{0}}r^{9}}$, $\mathcal{L} = -\frac{q^{2}}{4r^{4}} - \frac{q^{3}}{r^{5}} + \frac{ 5q^{5} }{14r^{7}} - \left(\!\!\frac{3q^{5} }{16m_{_{0}}r^{6}} - \frac{5q^{8} }{ 24m_{_{0}}r^{9} }\!\!\right)\!\ln(\beta)$, 
admits the AF-SSS magnetic massless black hole solution given by $-g_{tt}=1/g_{rr}=1-q^{3}/r^{3}$ for 
arbitrary values of the parameters ($m_{_{0}}$, $\beta$).
Hence, we conclude that the divergent terms in (\ref{regularizacion}), i.e., the terms proportional to $1/m_{0}$, cancel out in the ESTGB-NLED field equations.
Therefore, these divergent terms can be removed of the theory. Thus, the metric (\ref{NewBH}), with $m=0$, is a purely magnetic solution of the ESTGB-NLED model determined by;  
\begin{equation}
\boldsymbol{f}(\phi)=-\frac{q^{2}}{48\phi^{3}}=-\frac{r^{3}}{48q},\hskip.5cm \mathscr{U}(\phi) = \frac{9\phi^{7}}{56q^{2}} = \frac{9q^{5}}{56r^{7}}, \hskip.5cm \mathcal{L}(\mathcal{F}) = -\frac{\mathcal{F}}{4}\!-\!q^{\frac{1}{2}}\mathcal{F}^{\frac{5}{4}}\!+\!\frac{ 5q^{\frac{3}{2}}\mathcal{F}^{\frac{7}{4}} }{14} = -\frac{q^{2}}{4r^{4}}\!-\!\frac{q^{3}}{r^{5}}\!+\!\frac{ 5q^{5} }{14r^{7}}.    
\end{equation}
From here it is easy to see that $\mathcal{L}(\mathcal{F})$ satisfies the correspondence to Maxwell theory, i.e.,  $\mathcal{L} \rightarrow -\frac{\mathcal{F}}{4}$, $\mathcal{L}_{\mathcal{F}} \rightarrow -\frac{1}{4}$, as $\mathcal{F} \rightarrow 0$.

%%%%%%%%%%%%%%%%%%%%%%%%%%%%%%%%%%%%%%%%%%%%%%%%%%%%%%%%%%%%%%%%%%%%%%%%%%%%%%%%%%%%%%%%%%%%%%%%%%%%%%%%%%%%%

\section{Conclusion}

We have derived the first exact solution of
the ESTGB-NLED field equations
describing an asymptotically flat static and spherically symmetric black hole. 
The real scalar field 
constitutes a legitimate hair for the configuration, of the secondary type, since its charge is proportional to the magnetic charge of the black hole. 
Linear electrodynamics is recovered from the NLED presented here in the limit of sufficiently weak fields. 

By requiring that both the
effective
energy-momentum tensor and that of the NLED satisfy the weak energy condition, we have shown that only values of $q<0$ are allowed. 
In this case, 
the solution is akin to 
the Reissner-Nordstr\"om black hole, with the addition of the scalar hair: it can have two horizons, one, or none. 

In the limit of zero magnetic charge, the Schwarzschild solution is recovered. Another interesting limit is that achieved when $m=0$, leading to a a solution with magnetic charge (and scalar hair). 

It would be of interest to study the stability of these solutions
(a good starting point are the developments in \cite{Kanti1997}), as well as 
the corresponding thermodynamics (following \cite{Breton2015}).
We hope to return to these issues in a future publication.

\section*{Appendix}

In order to describe the limit cases of our solution, it is convenient to write the quantities $\boldsymbol{f}(\phi)$, $\mathscr{U}(\phi)$, and $\mathcal{L}(\mathcal{F})$ as functions of the radial coordinate. For the case with $\{m>0, q>0 \}$, theses are, 

\begin{eqnarray}
\boldsymbol{f}\!&=&\!-\frac{q^{2}}{32m}\!\!\left\{\!\frac{\sqrt{2}\hskip.05cm q^{2}}{\sqrt{qm}}\tan^{\!^{\!-1}}\!\!\!\left( \frac{\sqrt{2} \hskip.05cm mr }{ q\sqrt{qm}}\right) + \frac{r}{2}\ln\!\!\left[\!\!\left(\! \frac{ 2m\beta r^{2} }{q^{3}} + \beta\!\right)^{\!\!2}\right] - 2r\!\right\}\!\!, \label{f_r_New}\\%%%%%%%%
&&\nonumber\\
\mathscr{U}\!\!&=&\!\frac{2^{\!^{\frac{9}{2}}}\!m^{\!^{\frac{7}{2}}} }{ 105q^{\!^{\frac{11}{2}}} }\!\!\left[\!\frac{\pi}{2}\!-\!\tan^{\!^{\!-1}}\!\!\!\left(\!\!\frac{\sqrt{2}mr}{ q\sqrt{qm}}\!\!\right)  \!\right]\!+\!\frac{q^{2}}{4r^{5}}\!\!\left[\!\frac{3m}{10}\!+\!\frac{5q^{3}}{7r^{2}}\!+\! \frac{7q^{6}}{24mr^{4}}\!\right]\!\!\ln\!\!\left[\!\!\left(\!\frac{2m\beta r^{2}}{q^{3}}\!+\!\beta\!\right)^{\!\!2} \right]\!\!-\!\frac{1}{3qr}\!\!\left[\!\frac{16m^{3}}{35q^{3}}\!\!-\!\frac{8m^{2}}{105r^{2}} \!+\!\frac{31mq^{3}}{70r^{4}}\!+\!\frac{11q^{6}}{28r^{6}}\!\right]\!\!, \label{UNewr}\\ %%%%%%%%%%%%%%%%%%%%%%%%%%%%%%%%%%%%%%%%%%%%%%%%%%%%%%%%%%%%%%%%
\mathcal{L}\!&=&\!\frac{q^{2}}{8r^{4}}\!+\!\frac{8m^{3}}{105q^{4} r }\!-\!\frac{4m^{2}}{315q r^{3}}\!-\!\left(\!\frac{37m}{210}\!+\!q\!\right)\!\frac{ q^{2} }{ r^{5} } \!-\!\frac{ 5q^{5} }{84r^{7}}\!+\!\frac{ 2^{\!^{\frac{7}{2}}} m^{ \!^{\frac{7}{2}} } }{ 105 q^{\frac{11}{2}}  }\!\!\left[\!\tan^{\!^{\!-1}}\!\!\!\left(\!\!\frac{\sqrt{2} mr }{ q\sqrt{qm} }\!\!\right)\!- \frac{\pi}{2} \right] \nonumber \\
&&\hskip6.9cm -\frac{1}{2}\!\left(\!\frac{q^{2}}{8r^{4}}\!-\!\frac{ 3mq^{2} }{ 10r^{5} }\!+\!\frac{3 q^{5} }{16mr^{6}}\!-\!\frac{4 q^{5} }{7r^{7}}\!-\!\frac{ 5 q^{8} }{ 24mr^{9} }\!\right)\!\ln\!\!\left[\!\!\left(\!\frac{2m\beta r^{2}}{q^{3}}\!+\!\beta\!\right)^{\!\!2}\right]\!\!.\label{LNewr}
\end{eqnarray}
Then, by using (\ref{magnetica}), (\ref{NewBH}), (\ref{f_r_New}), (\ref{UNewr}), (\ref{LNewr}), $\dot{\boldsymbol{f}} = \frac{\boldsymbol{f}'}{\phi'}$, $\ddot{\boldsymbol{f}} =\frac{1}{\phi'} \left(\frac{\boldsymbol{f}'}{\phi'}\right)\!\!^{^{'}}$ and $\mathcal{L}_{\mathcal{F}} = \frac{\mathcal{L}'}{\mathcal{F}'}$, one finds that the field equations (\ref{Eqt}), (\ref{Eqr}), (\ref{Eqte}), and (\ref{phi2}), are satisfied. \\    
%%%%%%%%%%%%%%%%%%%%%%%%%
For the case with $\{m>0, q<0 \}$, rename $q=-Q$ (being $Q>0$), theses are
\begin{eqnarray}
\boldsymbol{f}\!&=&\!-\frac{Q^{2}}{32m}\!\!\left\{\!\frac{ Q^{2} }{2\sqrt{2Qm}}\!\ln\!\!\left[\!\!\left( \frac{ 1 + \frac{\sqrt{2} \hskip.05cm mr }{ Q\sqrt{Qm}} }{ 1 - \frac{\sqrt{2} \hskip.05cm mr }{ Q\sqrt{Qm}} }\!\!\right)^{\!\!\!2}\!\right] + \frac{r}{2}\ln\!\!\left[\!\!\left(\! \frac{ 2m\beta r^{2} }{Q^{3}} - \beta\!\right)^{\!\!2}\right] - 2r\!\right\}\!\!, \label{f_r_NewQ}\\%%%%%%%%
&&\nonumber\\
\mathscr{U}\!\!&=&\!\!\frac{2^{\!^{\frac{5}{2}}}\!m^{\!^{\frac{7}{2}}} }{ 105Q^{\!^{\frac{11}{2}}} }\!\ln\!\!\!\left[\!\!\left(\!\!\frac{1\!+\!\frac{\sqrt{2} \hskip.05cm mr }{ Q\!\sqrt{Qm}} }{ 1 \!-\! \frac{\sqrt{2} \hskip.05cm mr }{ Q\!\sqrt{Qm}} }\!\!\right)^{\!\!\!2}\!\right]\!\!+\!\frac{Q^{2}}{4r^{5}}\!\!\left[\!\frac{3m}{10}\!-\!\frac{5Q^{3}}{7r^{2}}\!+\! \frac{7Q^{6}}{24mr^{4}}\!\right]\!\!\ln\!\!\left[\!\!\left(\!\frac{2m\beta r^{2}}{Q^{3}}\!-\!\beta\!\right)^{\!\!\!2} \right]\!\!-\!\frac{1}{3Qr}\!\!\left[\!\frac{16m^{3}}{35Q^{3}}\!\!+\!\frac{8m^{2}}{105r^{2}} \!+\!\frac{31mQ^{3}}{70r^{4}}\!-\!\frac{11Q^{6}}{28r^{6}}\!\right]\!\!, \label{UNewrQ}\\ %%%%%%%%%%%%%%%%%%%%%%%%%%%%%%%%%%%%%%%%%%%%%%%%%%%%%%%%%%%%%%%%
\mathcal{L}\!&=&\!\frac{Q^{2}}{8r^{4}}\!+\!\frac{8m^{3}}{105Q^{4} r }\!+\!\frac{4m^{2}}{315Q r^{3}}\!-\!\left(\!\frac{37m}{210}\!-\!Q\!\right)\!\frac{ Q^{2} }{ r^{5} } \!+\!\frac{ 5Q^{5} }{84r^{7}}\!-\!\frac{ 2^{\!^{\frac{3}{2}}} m^{ \!^{\frac{7}{2}} } }{ 105 Q^{\frac{11}{2}}  }\!\ln\!\!\!\left[\!\!\left(\!\!\frac{1\!+\!\frac{\sqrt{2} \hskip.05cm mr }{ Q\!\sqrt{Qm}} }{ 1 \!-\! \frac{\sqrt{2} \hskip.05cm mr }{ Q\!\sqrt{Qm}} }\!\!\right)^{\!\!\!2}\!\right] \nonumber \\
&&\hskip6.5cm -\frac{1}{2}\!\left(\!\frac{Q^{2}}{8r^{4}}\!-\!\frac{ 3mQ^{2} }{ 10r^{5} }\!-\!\frac{3 Q^{5} }{16mr^{6}}\!+\!\frac{4Q^{5} }{7r^{7}}\!-\!\frac{ 5 Q^{8} }{ 24mr^{9} }\!\right)\!\ln\!\!\left[\!\!\left(\!\frac{2m\beta r^{2}}{Q^{3}}\!-\!\beta\!\right)^{\!\!2}\right]\!\!.\label{LNewrQ}
\end{eqnarray}
While for $\{m<0, q>0 \}$, rename $m=-M$ (being $M>0$), theses are
\begin{eqnarray}
\boldsymbol{f}\!&=&\!\frac{q^{2}}{32M}\!\!\left\{\!\frac{ q^{2} }{2\sqrt{2qM}}\!\ln\!\!\left[\!\!\left( \frac{ 1 + \frac{\sqrt{2} \hskip.05cm Mr }{ q\sqrt{qM}} }{ 1 - \frac{\sqrt{2} \hskip.05cm Mr }{ q\sqrt{qM}} }\!\!\right)^{\!\!\!2}\!\right] + \frac{r}{2}\ln\!\!\left[\!\!\left(\! \frac{ 2M\beta r^{2} }{q^{3}} - \beta\!\right)^{\!\!2}\right] - 2r\!\right\}\!\!, \label{f_r_NewM}\\%%%%%%%%
&&\nonumber\\
\mathscr{U}\!\!&=&\!\!\frac{2^{\!^{\frac{5}{2}}}\!M^{\!^{\frac{7}{2}}} }{ 105q^{\!^{\frac{11}{2}}} }\!\ln\!\!\!\left[\!\!\left(\!\!\frac{1\!-\!\frac{\sqrt{2} \hskip.05cm Mr }{ q\sqrt{qM}} }{ 1 \!+\! \frac{\sqrt{2} \hskip.05cm Mr }{ q\sqrt{qM}} }\!\!\right)^{\!\!\!2}\!\right]\!\!-\!\frac{q^{2}}{4r^{5}}\!\!\left[\!\frac{3M}{10}\!-\!\frac{5q^{3}}{7r^{2}}\!+\! \frac{7q^{6}}{24Mr^{4}}\!\right]\!\!\ln\!\!\left[\!\!\left(\!\frac{2M\beta r^{2}}{q^{3}}\!-\!\beta\!\right)^{\!\!\!2} \right]\!\!+\!\frac{1}{3qr}\!\!\left[\!\frac{16M^{3}}{35q^{3}}\!\!+\!\frac{8M^{2}}{105r^{2}} \!+\!\frac{31Mq^{3}}{70r^{4}}\!-\!\frac{11q^{6}}{28r^{6}}\!\right]\!\!, \label{UNewrM}\\ %%%%%%%%%%%%%%%%%%%%%%%%%%%%%%%%%%%%%%%%%%%%%%%%%%%%%%%%%%%%%%%%
\mathcal{L}\!&=&\!\frac{q^{2}}{8r^{4}}\!-\!\frac{8M^{3}}{105q^{4} r }\!-\!\frac{4M^{2}}{315q r^{3}}\!+\!\left(\!\frac{37M}{210}\!-\!q\!\right)\!\frac{ q^{2} }{ r^{5} } \!-\!\frac{ 5q^{5} }{84r^{7}}\!+\!\frac{ 2^{\!^{\frac{3}{2}}} M^{ \!^{\frac{7}{2}} } }{ 105 q^{\frac{11}{2}}  }\!\ln\!\!\!\left[\!\!\left(\!\!\frac{1\!+\!\frac{\sqrt{2} \hskip.05cm Mr }{ q\sqrt{qM}} }{ 1 \!-\! \frac{\sqrt{2} \hskip.05cm Mr }{ q\sqrt{qM}} }\!\!\right)^{\!\!\!2}\!\right] \nonumber \\
&&\hskip6.5cm -\frac{1}{2}\!\left(\!\frac{q^{2}}{8r^{4}}\!+\!\frac{ 3Mq^{2} }{ 10r^{5} }\!-\!\frac{3 q^{5} }{16Mr^{6}}\!-\!\frac{4q^{5} }{7r^{7}}\!+\!\frac{ 5 q^{8} }{ 24Mr^{9} }\!\right)\!\ln\!\!\left[\!\!\left(\!\frac{2M\beta r^{2}}{q^{3}}\!-\!\beta\!\right)^{\!\!2}\right]\!\!.\label{LNewrM}
\end{eqnarray}

\end{document}